\documentclass[12pt] {article}
\usepackage{psfrag}
\usepackage{graphicx}
\usepackage{latexsym,amsfonts}
\usepackage{amssymb}
\begin{document}
\setcounter{page}{1}
\def\theequation{\arabic{section}.\arabic{equation}}
\def\theequation{\thesection.\arabic{equation}}
\setcounter{section}{0}

\title{Phenomenological model of \\ the Kaonic
  Nuclear Cluster $K^-pp$ in the ground state}

\author{A. N. Ivanov$^{1,3}$\,\thanks{E--mail:
    ivanov@kph.tuwien.ac.at, Tel.: +43--1--58801--14261, Fax:
    +43--1--58801--14299}~\thanks{Permanent Address: State Polytechnic
    University, Department of Nuclear Physics, 195251 St. Petersburg,
    Russian Federation}, P.
  Kienle$^{2,3}$, J.  Marton$^3$, E.
  Widmann$^3$\,\thanks{http://www.oeaw.ac.at/smi}}

\date{\today}

\maketitle

\vspace{-0.5in}

\begin{center} 
{\it $^1$Atominstitut der \"Osterreichischen
    Universit\"aten, Technische Universit\"at Wien, Wiedner
    Hauptstrasse 8-10, A-1040 Wien, \"Osterreich, \\
    $^2$Physik Department, Technische Universit\"at M\"unchen,
    D--85748 Garching, Germany,\\ $^3$Stefan Meyer Institut f\"ur
    subatomare Physik, \"Osterreichische Akademie der Wissenschaften,
    Boltzmanngasse 3, A-1090, Wien, \"Osterreich}
\end{center}

\begin{center}
\begin{abstract}
  A phenomenological model is proposed for the analysis of the
  properties of the Kaonic Nuclear Cluster (KNC) $K^-pp$ (or
  ${^2_{\bar{K}}}{\rm H}$) in its ground state.  Inside the KNC
  ${^2_{\bar{K}}}{\rm H}$ we describe the relative motion of the
  protons in the $pp$ pair and the motion of the $K^-$--meson relative
  to the $pp$ pair in terms of wave functions of 3--dimensional
  harmonic oscillators.  The interaction strength is parameterised in
  our model by the frequency of the longitudinal oscillation of the
  $K^-$\,--\,meson relative to the $pp$ pair.  This parameter was
  determined with the binding energy and width of the strange baryon
  $\Lambda(1405)$ by assuming that it is a bound state of the $K^-p$
  pair. In terms of this interaction strength we calculate the binding
  energy $\epsilon_{{^2_{\bar{K}}}{\rm H}} = -\,118\,{\rm MeV}$, the
  partial widths of non\,--\,pionic decay channels and the total
  width, $\Gamma^{(\rm non-pion.)}_{{^2_{\bar{K}}}{\rm H}} = 58\,{\rm
    MeV}$. This agrees well with the experimental data by the FINUDA
  Collaboration (PRL {\bf 94}, 212303 (2005)):
  $\epsilon^{\exp}_{{^2_{\bar{K}}}{\rm H}} = -\,115^{+6}_{-5}\,{\rm
    MeV}$ and $\Gamma^{(\exp)}_{{^2_{\bar{K}}}{\rm H}} =
  (67^{+14}_{-11})\,{\rm MeV}$. The discrepancy with the results,
  obtained by Akaishi and Yamazaki  within the potential model approach, 
is discussed.
\end{abstract}  
\end{center}

\newpage

\section{Introduction}
\setcounter{equation}{0}

In this paper we propose a phenomenological model for the description
of the most simple Kaonic Nuclear Cluster (KNC) $K^-pp$, which we
denote ${^2_{\bar{K}}}{\rm H}$ \cite{TDB1}. For the first time the KNC
${^2_{\bar{K}}}{\rm H}$ has been investigated by Akaishi and Yamazaki
\cite{TDB1} within a non--relativistic Potential Model approach
\cite{TDB1}. They have predicted the following parameters of the KNC
${^2_{\bar{K}}}{\rm H}$: 1) the binding energy
$\epsilon_{{^2_{\bar{K}}}{\rm H}} = -\,48\,{\rm MeV}$, 2) the width
$\Gamma^{\rm non-pion.}_{{^2_{\bar{K}}}{\rm H}} \simeq 12\,{\rm MeV}$
of the non\,--\,pionic two\,--\,body decays ${^2_{\bar{K}}}{\rm H} \to
NY$, where $NY = p\Lambda^0, p\Sigma^0$ and $n\Sigma^+$ and 3) the
total width $\Gamma_{{^2_{\bar{K}}}{\rm H}} = 60\,{\rm MeV}$ by
including the open pionic channels \cite{TDB1,PK05}.

Recently the FINUDA Collaboration has reported \cite{FINUDA} results of
the analysis of the $p\Lambda^0$\,--\,correlations from
$K^-$\,--\,mesons stopped in light nuclear targets. They have been
interpreted as the formation of the bound $K^-pp$ system with a
binding energy of about $115\,{\rm MeV}$ and a total width of about
$70\,{\rm MeV}$.

The main aim of this paper is to formulate a phenomenological
  model for the KNC ${^2_{\bar{K}}}{\rm H}$ enabling to explain the
  experimental data by the FINUDA Collaboration \cite{FINUDA}.  For
  this we use the phenomenological approach for the description of
  kaonic atoms and strong low\,--\,energy $\bar{K}N$ and $\bar{K}NN$
  interactions, which has been recently developed in \cite{IV3}.

  The paper is organised as follows. In Section 2, following the
  analysis of the energy level shift and width of the ground state of
  kaonic deuterium \cite{IV3}, we adduce the general formulas for the
  binding energy and the width of the KNC ${^2_{\bar{K}}}{\rm H}$. In
  Section 3 we propose a phenomenological model for the calculation of
  the wave functions of the KNC ${^2_{\bar{K}}}{\rm H}$ with the $pp$
  pair in the ${^1}{\rm S}_0$ state. We draw an analogy with a
  symmetrical linear triatomic molecule. This allows to describe in
  terms of oscillator wave functions in the momentum representation
  the relative motion of the protons in the $pp$ pair and the motion
  of $K^-$\,--\,meson relative to the $pp$ pair inside the KNC
  ${^2_{\bar{K}}}{\rm H}$.  In our model the interaction strength is
  parameterised using the frequency $\omega$ of the longitudinal
  oscillation of the $K^-$ meson relative to the $pp$ pair inside the
  KNC ${^2_{\bar{K}}}{\rm H}$. In terms of $\omega$ we define the
  binding energy and the partial widths of non\,--\,pionic decays of
  the KNC ${^2_{\bar{K}}}{\rm H}$. In Section 4 we fix $\omega =
  64\,{\rm MeV}$ in terms of the binding energy and width of the
  strange baryon $\Lambda(1405)$ \cite{PDG04}.  We assume that 1) the
  strange baryon $\Lambda(1405)$ is a bound state of the $K^-p$ pair
  described by an oscillator wave function and 2) the linear restoring
  force has the same {\it stiffness} as in the KNC ${^2_{\bar{K}}}{\rm
    H}$.  In Section 5 we calculate the binding energy of the KNC
  ${^2_{\bar{K}}}{\rm H}$: $\epsilon_{{^2_{\bar{K}}}{\rm H}} =
  -\,118\,{\rm MeV}$.  In Sections 6 we calculate the partial widths
  of the non\,--\,pionic decay channels ${^2_{\bar{K}}}{\rm H} \to p
  \Lambda^0$, ${^2_{\bar{K}}}{\rm H} \to p \Sigma^0$ and
  ${^2_{\bar{K}}}{\rm H} \to n \Sigma^+$ and the total width
  $\Gamma_{{^2_{\bar{K}}}{\rm H}} = 58\,{\rm MeV}$. These values are
  in agreement with the experimental data by the FINUDA Collaboration
  \cite{FINUDA}. We predict that the dominant decay channel of KNC
  ${^2_{\bar{K}}}{\rm H}$ is ${^2_{\bar{K}}}{\rm H} \to p \Lambda^0$,
  whereas the decay channel ${^2_{\bar{K}}}{\rm H} \to p \Sigma^0$ is
  suppressed as $I_{p\Lambda^0}:I_{p\Sigma^0} = 17 : 1$, where
  $I_{p\Lambda^0}$ and $I_{p\Sigma^0}$ are the intensities of the
  invariant\,--\,mass spectra of the $p\Lambda^0$ and the $p\Sigma^0$
  pairs.  In the Conclusion we discuss the obtained results.  In the
  Discussion we show that the binding energy and the width of the KNC
  ${^2_{\bar{K}}}{\rm H}$ depends on the structure of the $K^-pp$
  system. We calculate the binding energy and the width following the
  structure of the KNC ${^2_{\bar{K}}}{\rm H} = (K^-p)_{I = 0}\otimes
  p$, induced due to the $\Lambda(1405)\otimes p$ doorway dominance
  proposed by Akaishi and Yamazaki \cite{Yamazaki}. We show that our
  oscillator model describes the KNC ${^2_{\bar{K}}}{\rm H}$ with the
  structure $(K^-p)_{I = 0}\otimes p$ in quantitative agreement with
  the results, obtained by Akaishi and Yamazaki within the potential
  model approach \cite{TDB1,Yamazaki}. In Appendix A we represent the
  detailed expressions for the binding energy and the width of the KNC
  ${^2_{\bar{K}}}{\rm H}$ in terms of the wave functions of the KNC
  ${^2_{\bar{K}}}{\rm H}$ and the S\,--\,wave amplitude of elastic
  $K^-pp$ scattering. In Appendix B we adduce the effective chiral
  Lagrangian invariant under non--linear transformations of chiral
  $SU(3)\times SU(3)$ symmetry with derivatives meson--baryon
  couplings.  In Appendix C we calculate the pseudoscalar meson
  propagators weighted with the wave function of the KNC
  ${^2_{\bar{K}}}{\rm H}$.  These results are applied to the
  calculation of the partial widths of the KNC ${^2_{\bar{K}}}{\rm
    H}$.

\section{Definition of the binding energy and width of the
  ${^2_{\bar{K}}}{\rm H}$}
\setcounter{equation}{0}

In our analysis of the KNC ${^2_{\bar{K}}}{\rm H}$ we propose to
describe the binding energy $\epsilon_{{^2_{\bar{K}}}{\rm H}}$ and the
width $\Gamma_{{^2_{\bar{K}}}{\rm H}}$ of the KNC ${^2_{\bar{K}}}{\rm
  H}$ by analogy with the energy level shift and width of the ground
state of kaonic deuterium \cite{IV3}. In this case
$\epsilon_{{^2_{\bar{K}}}{\rm H}}$ and $\Gamma_{{^2_{\bar{K}}}{\rm
    H}}$ can be defined in terms of the S\,--\,wave amplitude $M(K^-pp
\to K^-pp)$ of elastic $K^-pp$ scattering with the $pp$ pair in the
${^1}{\rm S}_0$ state weighted with the wave functions of the KNC
${^2_{\bar{K}}}{\rm H}$. This gives
\begin{eqnarray}\label{label2.1}
  \hspace{-0.3in}&&  -\,\epsilon_{{^2_{\bar{K}}}{\rm H}} +\,i\,
  \frac{\Gamma_{{^2_{\bar{K}}}{\rm H}}}{2} = \int d\tau\,\Phi^*_{{^2_{\bar{K}}}{\rm H}}\,
M(K^-pp \to K^-pp)\Phi_{{^2_{\bar{K}}}{\rm H}},
\end{eqnarray}
where $\Phi_{{^2_{\bar{K}}}{\rm H}}$ is the wave function of the KNC
${^2_{\bar{K}}}{\rm H}$ in the momentum representation and $d\tau$ is
the phase volume element of the interacting particles.

The width $\Gamma_{{^2_{\bar{K}}}{\rm H}}$ is the sum of the
non--pionic two\,--\,body decays $K^-pp \to NY$, where $NY =
p\Lambda^0, p\Sigma^0$ and $n\Sigma^+$, and pionic three\,--\,body
decays $K^-pp \to NY\pi$, where $NY\pi = \Sigma N \pi$ and $\Lambda^0
N\pi$. It is given by
\begin{eqnarray}\label{label2.2}
  \Gamma_{{^2_{\bar{K}}}{\rm H}} = 
  \sum_{NY}\Gamma({^2_{\bar{K}}}{\rm H} \to NY) + \sum_{NY\pi}
  \Gamma({^2_{\bar{K}}}{\rm H} \to NY\pi ).
\end{eqnarray}
The formulas (\ref{label2.1}) and (\ref{label2.2}) are presented in
detail in Appendix A.

\section{Model for the Kaonic Nuclear Cluster 
${^2_{\bar{K}}}{\rm H}$}
\setcounter{equation}{0}

The calculation of the S\,--\,wave amplitude $M(K^-pp \to K^-pp)$ is
carried out below by using Effective Chiral Lagrangians (see Appendix
B).  For the description of the wave function
$\Phi_{{^2_{\bar{K}}}{\rm H}}$ we use an analogy between a symmetrical
linear triatomic molecule and ${^2_{\bar{K}}}{\rm H}$ shown in Fig.\,1.
\begin{figure}
\centering \psfrag{K-}{$K^-$} 
\psfrag{p}{$p$}
\psfrag{a}{$\vec{q}$}
\psfrag{b}{$-\,\frac{1}{2}\,\vec{q}$}
\psfrag{c}{$+\,\vec{Q}$}
\psfrag{d}{$-\,\vec{Q}$}
\psfrag{(A)}{$(1a)$}
\psfrag{(B)}{$(1b)$}
\psfrag{(C)}{$(1c)$}
\includegraphics[height=0.37\textheight]{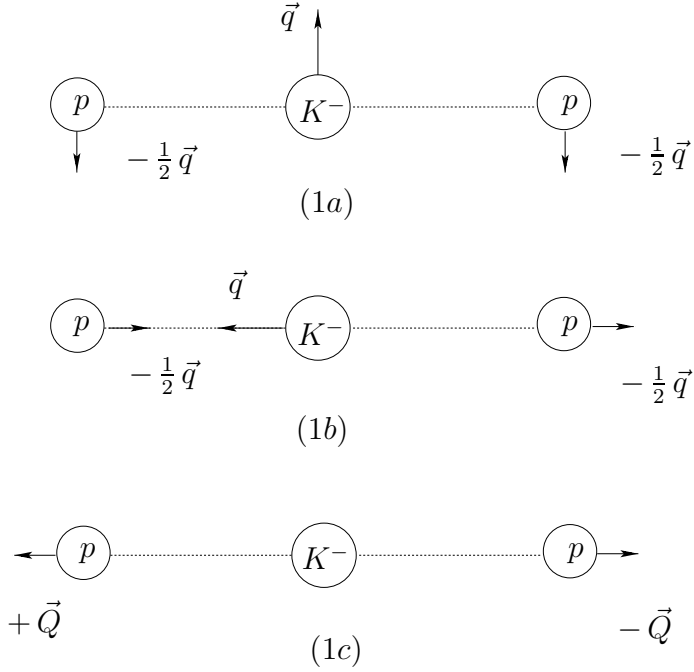}
\caption{The Kaonic Nuclear Cluster ${^2_{\bar{K}}}{\rm H}$ in the
  ground state.}
\end{figure}

A symmetrical linear triatomic molecule ${^2_{\bar{K}}}{\rm H}$ as
system of coupled harmonic oscillators should possess one
transverse vibrational mode, depicted in Fig.\,1a, and two longitudinal
modes, depicted in Fig.\,1b and Fig.\,1c. The frequencies of these modes
are defined by \cite{HG59}
\begin{eqnarray}\label{label3.1}
  \omega_{\,\perp} = 2\,\sqrt{\frac{k_{\,\perp}}{\mu}}\quad,
\quad \omega_{\,\parallel} = \sqrt{\frac{2 k_{\,\parallel}}{\mu}}\quad,
\quad \Omega_{\,\parallel} = 
  \sqrt{\frac{k_{\,\parallel}}{2 \mu_N}} = \omega_{\,\parallel}\sqrt{\frac{\mu}{2m_N}},
\end{eqnarray} 
where $\mu = 2m_Km_N/(m_K + 2 m_N) = 391\,{\rm MeV}$ is the reduced
mass of the KNC ${^2_{\bar{K}}}{\rm H}$, calculated for $m_K =
494\,{\rm MeV}$ and $m_p = m_N = 940\,{\rm MeV}$, and $\mu_N = m_N/2$
is the reduced mass of the $pp$ pair, and  $k_{\,\perp}$ and
$k_{\,\parallel}$ are the {\it stiffnesses} for linear restoring
forces acting on the transverse and longitudinal degrees of freedom of
the symmetrical linear triatomic molecule ${^2_{\bar{K}}}{\rm H}$.

The frequencies (\ref{label3.1}) define independent degrees of freedom
of the $K^-pp$ system. We identify the frequency
$\Omega_{\,\parallel}$ with the relative motion of the protons in the
$pp$ pair inside the KNC ${^2_{\bar{K}}}{\rm H}$. Such a motion can be
described by the wave function $\Phi_{pp}(\vec{Q}\,)$ \cite{LL65}:
\begin{eqnarray}\label{label3.2}
  \Phi_{pp}(\vec{Q}\,) = \frac{(4\pi)^{3/4}}{(\mu^3_N
  \Omega^3_{\,\parallel})^{1/4}}\,
  \exp\Big(-\,\frac{\vec{Q}^{\,2}}{2\mu_N \Omega_{\,\parallel}}\Big) =
  \frac{(8\pi)^{3/4}}{(m^3_N \Omega^3_{\,\parallel})^{1/4}}\,
  \exp\Big(-\,\frac{\vec{Q}^{\,2}}{m_N \Omega_{\,\parallel}}\Big).
\end{eqnarray} 
The wave function (\ref{label3.2}) is spherical symmetric, hence, 
corresponding to a spherical symmetric 3\,--\,dimensional harmonic
oscillator. The contribution of these degrees of freedom to the
binding energy of the KNC ${^2_{\bar{K}}}{\rm H}$ is equal to
$3\,\Omega_{\,\parallel}/2$.

The frequencies $\omega_{\,\perp}$ and $\omega_{\,\parallel}$ we
identify with the degrees of freedom of a motion of the kaon with a
reduced mass $\mu$.  This motion one can describe by the wave function
$\Phi_{K^-pp}(\vec{q}\,)$ \cite{LL65}:
\begin{eqnarray}\label{label3.3}
  \Phi_{K^-pp}(\vec{q}\,) = \frac{(4\pi)^{3/4}}{(\mu^3\omega_{\,\parallel}
    \omega^2_{\,\perp})^{1/4}}\,\exp\Big(-\,
  \frac{q^2_{\,\parallel}}{2\mu \omega_{\,\parallel}} - 
  \,\frac{\vec{q}^{\;2}_{\,\perp}}{2\mu \omega_{\,\perp}}\Big).
\end{eqnarray} 
This is a wave function of a 3\,--\,dimensional asymmetric harmonic
oscillator. The contribution of these degrees of freedom to the
binding energy of the KNC ${^2_{\bar{K}}}{\rm H}$ amounts to
$(\omega_{\,\parallel} + 2 \omega_{\,\perp})/2$. The wave functions
$\Phi_{pp}(\vec{Q}\,)$ and $\Phi_{K^-pp}(\vec{q}\,)$ are normalised to
unity.  Summing up the contributions, the binding energy of the KNC
${^2_{\bar{K}}}{\rm H}$ is defined by
\begin{eqnarray}\label{label3.4}
  \epsilon_{{^2_{\bar{K}}}{\rm H}} = U_0 + \frac{1}{2}\,(3\,\Omega_{\,\parallel}  + 
2\,\omega_{\,\perp} + \omega_{\,\parallel} ),
\end{eqnarray} 
where $U_0$ is a depth of the potential well, providing a harmonic
motion of the constituents of the KNC ${^2_{\bar{K}}}{\rm H}$. 

We assume that the {\it stiffnesses} for the longitudinal and
transverse motion of the vibrational modes of the KNC
${^2_{\bar{K}}}{\rm H}$ are equal $k_{\,\parallel} = k_{\,\perp} = k$
and the frequencies $\Omega_{\,\parallel}$ and $\omega_{\,\perp}$
depend on $\omega = \sqrt{2k/\mu}$ only. This means that the
interaction is assumed to be isotropic in the nuclear molecule.  The
averaged values of the squared momenta of the motion of the
$K^-$\,--\,meson relative to the $pp$ pair and the relative motion of
the protons in the $pp$ pair in the KNC ${^2_{\bar{K}}}{\rm H}$ are
equal to $q^2_0 = \langle q^2_{\,\parallel} + q^2_{\,\perp}\rangle =
(1 + 2 \sqrt{2})\,\mu \omega/2 = 748.5\,\omega \,{\rm MeV}^2$ and
$Q^2_0 = \langle Q^2\rangle = 3\omega\sqrt{\mu m_N}/4\sqrt{2} =
321.5\,\omega\,{\rm MeV}^2$, respectively. The momenta $q_0$ and $Q_0$
characterise also the spatial extension of the cluster
${^2_{\bar{K}}}{\rm H}$.

\section{Determination of the interaction strength $\omega$}
\setcounter{equation}{0}

In order to fix the interaction strength parameter $\omega$ we have to
investigate a system simpler than the KNC ${^2_{\bar{K}}}{\rm H}$, the
mass and width of which are well measured and which can be treated as
a quasi--bound state. One of the most obvious candidates for the
simplest KNC is the strange baryon $\Lambda(1405)$ with mass
$m_{\Lambda(1405)} = 1406 \pm 4\,{\rm MeV}$ and the width
$\Gamma_{\Lambda(1405)} = 50\pm 2\,{\rm MeV}$ \cite{PDG04}.  According
to Akaishi and Yamazaki \cite{TDB1}, the strange baryon
$\Lambda(1405)$ can be treated as a bound state of the $K^-p$ pair or
${^1_{\bar{K}}}{\rm H}$ with isospin $I = 0$. The binding energy of
the KNC ${^1_{\bar{K}}}{\rm H}$ is equal to
$\epsilon_{{^1_{\bar{K}}}{\rm H}} = m_{{^1_{\bar{K}}}{\rm H}} - m_K -
m_N = -\,28\,{\rm MeV}$, where we have set $m_{{^1_{\bar{K}}}{\rm H}}
= 1406\,{\rm MeV}$ \cite{PDG04}.  The width of ${^1_{\bar{K}}}{\rm H}$
equal to $\Gamma_{{^1_{\bar{K}}}{\rm H}} = 50\,{\rm MeV}$ is defined
by the ${^1_{\bar{K}}}{\rm H} \to \pi \Sigma$ pionic decay channels
only.

The assumption that the strange baryon $\Lambda(1405)$ can be treated
as the KNC ${^1_{\bar{K}}}{\rm H}$ does not contradict the results of
the analysis of this state within the combination of Chiral
Perturbation Theory (ChPT) \cite{JG83} with non--perturbative
coupled--channels techniques \cite{WW95,WW98} (see also \cite{WW05}).
In such an approach the strange baryon $\Lambda(1405)$ has been
generated ``dynamically as an $I = 0$ $\bar{K}N$ quasi--bound state
and as a resonance in the $\pi \Sigma$ channel'' \cite{WW05}.

Assuming that the relative motion of the $K^-p$ pair is defined by a
linear restoring force with the {\it stiffness} equal to the {\it
  stiffness} in the KNC ${^2_{\bar{K}}}{\rm H}$, we can write the wave
function of the KNC ${^1_{\bar{K}}}{\rm H}$ in the momentum
representation:
\begin{eqnarray}\label{label4.1}
  \Phi_{{^1_{\bar{K}}}{\rm H}}(\vec{q}\,) =
  \Big(\frac{4\pi}{\bar{\mu}\Omega}\Big)^{3/4}\,
\exp\Big(-\,\frac{\vec{q}^{\;2}}{2\bar{\mu}\Omega}\Big),
\end{eqnarray} 
where the frequency $\Omega$ is related to $\omega$ as $\Omega =
\omega\sqrt{\mu/2\bar{\mu}}$. The binding energy and the width of the
KNC ${^1_{\bar{K}}}{\rm H}$ are  defined by
\begin{eqnarray}\label{label4.2}
  \hspace{-0.3in} -\epsilon_{{^1_{\!\bar{K}}}{\rm H}}
 + i\,\frac{\Gamma_{{^1_{\!\bar{K}}}{\rm H}}}{2} = 
  \frac{1}{4m_K m_p}\int\!\!\!\int \frac{d^3kd^3q}{(2\pi)^6}
  \Phi^{\dagger}_{{^1_{\!\bar{K}}}{\rm H}}(\vec{k}\,)\,
  M((K^-p)_{I = 0} \to
  (K^-p)_{I = 0})\,
  \Phi_{{^1_{\!\bar{K}}}{\rm H}}(\vec{q}\,),
\end{eqnarray} 
where $M(K^-p)_{I = 0} \to (K^-p)_{I = 0})$ is the S\,--\,wave amplitude
of elastic $K^-p$ scattering in the state with isospin $I = 0$.

 We calculate the S--wave amplitudes of $K^-p$ and $K^-pp$
  scattering within Chiral Lagrangian approach with the non--linear
  realization of chiral $SU(3)\times SU(3)$ symmetry \cite{BWL68} (see
  Appendix B) to leading order in the large $N_C$ expansion, where
  $N_C$ is the number of quark colour degrees of freedom in
  multi--colour QCD \cite{EW79}, accepted for the analysis of
  meson--baryon interactions within ChPT \cite{JG83}--\cite{EK01}.
  Since in the multi--colour QCD masses of ground baryons are
  proportional to $N_C$ for $N_C \to \infty$ \cite{EW79}, the large
  $N_C$ expansion corresponds to the heavy--baryon limit, accepted in
  ChPT for the analysis of baryon exchanges \cite{JG83}--\cite{EK01}
  (see also \cite{AI1}--\cite{AI99}).

 The use of the large $N_C$ expansion allows to reduce the number
  of Feynman diagrams, contributing to the amplitudes of $K^ -p$ and
  $K^ -pp$ scattering, and to represent the final expression in terms
  of effective local Lagrangians. The former is important for the
  analysis of interactions of baryon pairs in the states with a
  certain angular momentum.

To leading order in the large $N_C$ expansion the real part of the
S\,--\,wave amplitude $M(K^-p)_{I = 0} \to (K^-p)_{I = 0})$ is defined
by the Weinberg\,--\,Tomozawa term \cite{WT66}--\cite{IV3a}. The
result, which can be obtained by the chiral Lagrangian ${\cal
  L}_{PPBB}(x)$ (see Appendix B Eq.({\rm B}.9)), is
\begin{eqnarray}\label{label4.3}
{\cal R}e
M((K^-p))_{I = 0} \to
(K^-p))_{I = 0}) = M^{WT}_{K^-p} = \frac{3 m_K m_N}{F^2_{\pi}},
\end{eqnarray} 
where $F_{\pi} = 92.4\,{\rm MeV}$ is the PCAC constant \cite{PDG04}.

 Since in the large $N_C$ expansion for $N_C \to \infty$ the
  baryon mass $m_N$ and the PCAC constant $F_{\pi}$ behave as $m_N =
  O( N_C)$ and $F_{\pi} =O(\sqrt{N_C})$ \cite{EW79}--\cite{EK01}, the
  contribution of the Weinberg--Tomozawa term is of order $O(1)$. The
  widths of resonances are of order $O(1/N_C)$ \cite{EW79}. As a
  result the coupling constant $g_{\Lambda^0(1405)}$ of the
  $\Lambda^0(1405)$ resonance with the $\pi \Sigma$ and $\bar{K}N$
  pairs is of order $g_{\Lambda^0(1405)} = O(1/N_C)$ \cite{IV3}.
  Hence, for $N_C \to \infty$ the contribution of the
  $\Lambda^0(1405)$ resonance to the amplitude $K^-p$ scattering is of
  order $O(1/N_C)$ \cite{IV3}: ${\cal R}e\,M^{\Lambda^0(1405)}_{K^ -p}
  = - g^2_{\Lambda^0(1405)}\,4m_K m_N/p^2_0 = O(1/N_C)$, where $p_0 =
  \sqrt{3 \bar{\mu}\Omega/2} = 150\,{\rm MeV}$. Hence, to leading
  order in the large $N_C$ expansion the contribution of the
  Weinberg--Tomozawa term dominates.

  This dominance retains also for $N_C = 3$. Indeed the contribution
  of the $\Lambda^0(1405)$ resonance, relative to the
  Weinberg--Tomozawa term, is equal to \cite{WW76}
  \begin{eqnarray}\label{label4.4}
    \Big|\frac{{\cal R}e\,M^{\Lambda^0(1405)}_{K^ -p}}{M^{WT}_{K^ -p}}\Big| = 
    \frac{2}{3}\,\frac{F^2_{\pi}}{m_K}\,\frac{( \sqrt{s} - 
      m_{\Lambda^0(1405)})g^2_{\Lambda^0(1405)}}{\displaystyle (m_{\Lambda^0(1405)} - 
      \sqrt{s}\,)^2 + \frac{\Gamma^2_{\Lambda^0(1405)}}{4}} = 0.12,
\end{eqnarray} 
where $\sqrt{s} = 1470\,{\rm MeV}$ is the energy of the $K^-p$ pair
with a relative momentum $p_0 = \sqrt{3 \bar{\mu}\Omega/2} = 150\,{\rm
  MeV}$ and $g^2_{\Lambda^0(1405)} = 0.90$ \cite{IV3}, defined for
$\Gamma_{\Lambda^0(1405)} = 53\,{\rm MeV}$ (see Eq.(\ref{label4.6})).

When vector mesons are included the effective chiral Lagrangian,
responsible for the Weinberg--Tomozawa term, can be obtained in the
one--vector--meson exchange approximation. In the non--linear
realization of chiral $SU(2)\times SU(2)$ symmetry this has been shown
by Weinberg \cite{SW68}. For the non--linear realization of chiral
$SU(3)\times SU(3)$ symmetry this can be shown by using the results of
the analysis of the vector meson interactions within ChPT carried out
by Ecker {\it et al.} \cite{JG85}. Using \cite{JG85} one can show that
relative to the Weinberg--Tomozawa term the contributions of
multi-vector--meson exchanges are of next--to--leading order in the
large $N_C$ and chiral expansion. 

The binding energy of the KNC ${^1_{\!\bar{K}}}{\rm H}$ is given by
\begin{eqnarray}\label{label4.5}
  -\,\epsilon_{{^1_{\!\bar{K}}}{\rm H}} = \frac{3}{4}\,\frac{1}{F^2_{\pi}}\Big\vert 
\int \frac{d^3q}{(2\pi)^3}\,\Phi_{{^1_{\!\bar{K}}}{\rm H}}(\vec{q}\,)\Big\vert^2 = 
\frac{3}{4}\,\frac{1}{F^2_{\pi}}\,
\Big(\frac{\omega}{\pi}\sqrt{\frac{\bar{\mu} \mu}{2}}\Big)^{3/2},
\end{eqnarray} 
where we have used the relation $\Omega =
\omega\sqrt{\mu/2\bar{\mu}}$.  In our approach the width of the KNC
${^1_{\!\bar{K}}}{\rm H}$ is related to the binding energy as
\begin{eqnarray}\label{label4.6}
 \Gamma_{{^1_{\!\bar{K}}}{\rm H}} = \frac{3}{4\pi}\,\frac{m_Km_N}{m_K +
  m_N + \epsilon_{{^1_{\!\bar{K}}}{\rm
      H}}}\,\frac{k_{\Sigma\pi}}{F^2_{\pi}}\,(-\,\epsilon_{{^1_{\!\bar{K}}}{\rm
    H}}),
\end{eqnarray} 
where $k_{\Sigma\pi} = 180\,{\rm MeV}$ is a relative momentum of the
$\Sigma\pi$ pair. Setting $\omega = 64\,{\rm MeV}$ we get the binding
energy $\epsilon_{{^1_{\!\bar{K}}}{\rm H}} = -\,32\,{\rm MeV}$ or the
mass of the bound state $m_{{^1_{\!\bar{K}}}{\rm H}} = 1402\,{\rm
  MeV}$ and $\Gamma_{{^1_{\!\bar{K}}}{\rm H}} = 53\,{\rm MeV}$, which
agree well with the mass and width of the $\Lambda^0(1405)$ resonance,
obtained by Dalitz and Deloff as the position of the pole on sheet II
of the $E$--plane $E^* - i\,\Gamma/2$ with $E^* = 1404.9\,{\rm MeV}$
and $\Gamma = 53.1\,{\rm MeV}$ \cite{RD91}. 

Since the $K^-p$ pair couples in the isospin--singlet state, in order
to understand how well our approximation describes the $K^-p$
scattering we can calculate the cross section for the reaction $K^- p
\to \Sigma^0\pi^0$. We get $\sigma(K^-p \to \Sigma^0\pi^0) = (38\pm 5)\,{\rm
  mb}$. This result agrees well with $\sigma(K^-p \to \Sigma^0\pi^0) =
(30\pm 4)\,{\rm mb}$ obtained in \cite{IV3}. We would like to
accentuate that the amplitude of the reaction $K^-p \to \Sigma^0\pi^0$
together with the amplitudes of the reactions $K^-p \to
\Sigma^{\pm}\pi^{\mp}$ and $K^-p\to \Lambda^0\pi^0$, calculated in
\cite{IV3}, describe well the experimental data on the energy level
width of the ground state of kaonic hydrogen by the DEAR Collaboration
(see \cite{IV3} and references therein).

\section{Binding energy of the ground state of KNC
  ${^2_{\bar{K}}}{\rm H}$}
\setcounter{equation}{0}

According to relation (\ref{label2.1}), the binding energy of the KNC
${^2_{\!\bar{K}}}{\rm H}$ is defined by the real part of the
S\,--\,wave amplitude $M(K^-pp \to K^-pp)$ of elastic $K^-pp$
scattering weighted with the wave functions of the KNC
${^2_{\bar{K}}}{\rm H}$ (see ({\rm A}.1)).

To leading order in the large $N_C$ and chiral expansion the
S\,--\,wave amplitude of elastic $K^-pp$ scattering is defined by the
contribution of the Weinberg\,--\,Tomozawa term and Feynman diagrams,
depicted in Fig.\,2.
\begin{figure}
  \centering \psfrag{K-}{$K^-$} \psfrag{p}{$p$} \psfrag{b}{$+ ~\ldots
    $}
\includegraphics[height= 0.15\textheight]{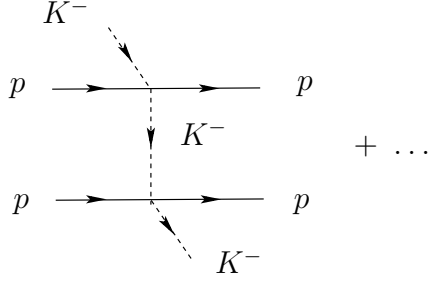}
\caption{The Feynman diagrams giving in the tree--approximation the
  main contribution to the S\,--\,wave amplitude of the reaction
  $K^-pp \to K^-pp$ to leading order in the large $N_C$ and chiral
  expansion.}
\end{figure}
These diagrams are defined by the chiral Lagrangian ${\cal L}_{PPBB}$
(see Appendix B). The result of the calculation of the diagrams in
Fig.\,2 can be represented by the effective local Lagrangian
\cite{AI1}
\begin{eqnarray}\label{label5.1}
  \hspace{-0.3in}M(K^-(pp)_{{^1}{\rm S}_0}\to 
  K^-(pp)_{{^1}{\rm S}_0}) = 
  - 2\,\langle C^{K^-(pp)_{{^1}{\rm S}_0}}_{K^-(pp)_{{^1}{\rm S}_0}}\rangle 
  [\bar{u}_p\gamma^5 u^c_p][\bar{u^c}_p\gamma^5 u_p],
\end{eqnarray}
The spinorial wave functions of the $pp$ pairs in the initial and the
final state are described by $[\bar{u}_p\gamma^5 u^c_p]$ $[\bar{u^c}_p
\gamma^5 u_p]$, respectively, where $u^c_p = C\bar{u}^T_p$ and
$\bar{u^c}_p = u^T_p C$ are charge conjugate Dirac bispinors.  The
effective coupling constant $\langle C^{K^-(pp)_{{^1}{\rm
      S}_0}}_{K^-(pp)_{{^1}{\rm S}_0}}\rangle$ is weighted with the
wave functions of the KNC ${^2_{\bar{K}}}{\rm H}$. It is equal to
\begin{eqnarray}\label{label5.2}
  \langle C^{K^-(pp)_{{^1}{\rm S}_0}}_{K^-(pp)_{{^1}{\rm S}_0}}\rangle = 
\frac{m^2_K}{F^4_{\pi}}\Big\langle \frac{1}{m^2_K - Q^2_K}\Big\rangle.
\end{eqnarray}
The correction $\langle \delta C^{K^-(pp)_{{^1}{\rm
      S}_0}}_{K^-(pp)_{{^1}{\rm S}_0}}\rangle$ to the effective
coupling constant $\langle C^{K^-(pp)_{{^1}{\rm
      S}_0}}_{K^-(pp)_{{^1}{\rm S}_0}}\rangle$ is defined by the
diagrams in Fig.\,3.
\begin{figure} 
\centering
  \psfrag{K-}{$K^-$} 
\psfrag{p}{$p$} 
\psfrag{Y}{$\Lambda^0,\Sigma^0$}
\psfrag{b}{$+ ~\ldots$}
\psfrag{a}{$+$}
\includegraphics[height= 0.16\textheight]{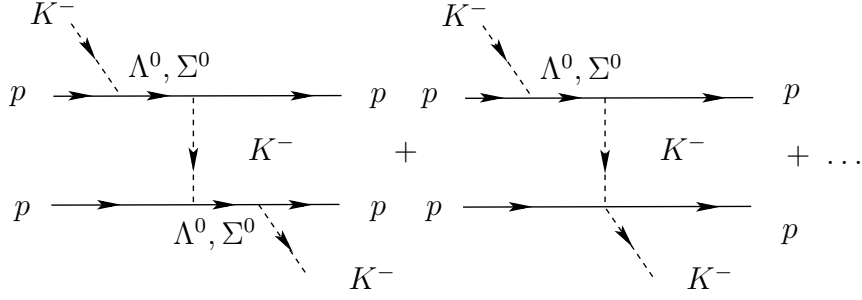}
\caption{The Feynman diagrams giving in the tree--approximation the
  main corrections to the S\,--\,wave amplitude of the reaction $K^-pp
  \to K^-pp$ defined by the diagrams in Fig.\,2, calculated to leading
  order in the large $N_C$ and chiral expansion.}
\end{figure}
This correction is of the next--to--leading order
in the large $N_C$ expansion and chiral expansion. Hence, we can drop
it. For $N_C = 3$ the effective coupling constant $\langle \delta
C^{K^-(pp)_{{^1}{\rm S}_0}}_{K^-(pp)_{{^1}{\rm S}_0}}\rangle$,
calculated with the chiral Lagrangians in Appendix B, is equal to
\begin{eqnarray}\label{label5.3}
  \hspace{-0.3in}\frac{\langle \delta C^{K^-(pp)_{{^1}{\rm
          S}_0}}_{K^-(pp)_{{^1}{\rm S}_0}}\rangle}{\langle 
    C^{K^-(pp)_{{^1}{\rm S}_0}}_{K^-(pp)_{{^1}{\rm S}_0}}\rangle} &=&-\,\frac{g^2_A}{4}\,
  \Big[\Big(\frac{3 - 2 \alpha_D}{\sqrt{3}}\Big)^2 
  + (2\alpha_D - 1)^2\Big]
  \frac{m_K}{2 m_N + m_K}\nonumber\\
  \hspace{-0.3in}&&+  \frac{g^4_A}{16}\,
  \Big[\Big(\frac{3 - 2 \alpha_D}{\sqrt{3}}\Big)^2  
  + (2\alpha_D - 1)^2\Big]^2\frac{m^2_K}{(2 m_N + m_K)^2 } = -\,0.08.
\end{eqnarray}
Hence the correction to the effective coupling constant $\langle
C^{K^-(pp)_{{^1}{\rm S}_0}}_{K^-(pp)_{{^1}{\rm S}_0}}\rangle$ makes up
$8\,\%$ only.  The contribution to the $K^-pp$ scattering, defined by
the effective coupling constant $\langle C^{K^-(pp)_{{^1}{\rm
      S}_0}}_{K^-(pp)_{{^1}{\rm S}_0}}\rangle$, is of the
next--to--leading order in the large $N_C$ and chiral expansion with
respect to the Weinberg--Tomozawa term. Since we keep only the leading
order contributions, so the contribution of the diagrams in Fig.\,2 can
be omitted. For $N_C = 3$ one can show that the contribution to the
binding energy of the KNC ${^2_{\bar{K}}}{\rm H}$ of the diagrams in
Fig.\,2 makes up about $5\,\%$. Thus, the binding energy of the KNC
${^2_{\bar{K}}}{\rm H}$ is defined by the Weinberg--Tomozawa term only
\begin{eqnarray}\label{label5.4}
  \hspace{-0.3in} -\,\epsilon_{{^2_{\bar{K}}}{\rm H}} = \frac{1}{F^2_{\pi}}\Big|
  \int \frac{d^3q }{(2\pi)^3}\,
  \Phi_{K^-pp}(\vec{q}\,)\Big|^2 = \sqrt{\frac{2}{\pi^3}}\,\frac{\mu^3}{F^2_{\pi}}\,
  \Big(\frac{\omega}{\mu}\Big)^{3/2} =  118\,{\rm MeV},
\end{eqnarray}
where we have set the $\omega = 64\,{\rm MeV}$ and used the momentum
integrals over the momenta $\vec{q}$ and $\vec{Q}$ 
\begin{eqnarray}\label{label5.5}
  \int \frac{d^3q }{(2\pi)^3}\,
  \Phi_{K^-pp}(\vec{q}\,) =\Big(\frac{2\mu^3\omega^3}{\pi^3}\Big)^{1/4},
\int \frac{d^3Q}{(2\pi)^3}\,\Phi_{pp}(\vec{Q}\,) =
  \Big(\frac{m^3_N\omega^3}{8\pi^3}\Big)^{1/4}\Big(\frac{\mu}{2m_N}\Big)^{3/8}.
\end{eqnarray}
The integrals over the momenta $\vec{q}$ and $\vec{Q}$ one needs also
for the calculation of the partial widths of the KNC
${^2_{\bar{K}}}{\rm H}$ decays.

We can calculate the contribution of the two\,--\,nucleon
absorption channels $K^-pp \to NY$. Using the results obtained in
Section 6 and in \cite{IV3} one can show that the contribution of the
two\,--\,nucleon absorption channels $K^-pp \to NY$ is about $1.5\,\%$
of the Weinberg--Tomozawa term.

\section{Width of the ground state of KNC
  ${^2_{\bar{K}}}{\rm H}$ }
\subsection{Decay ${^2_{\bar{K}}}{\rm H} \to p \Lambda^0$ }
\setcounter{equation}{0}

The amplitude of the reaction $K^-
pp \to p\Lambda^0$ is defined by
the Feynman diagrams in Fig.\,4. This gives
\begin{eqnarray}\label{label6.1}
  \hspace{-0.3in}&& M(K^-pp \to p\Lambda^0) =
  i\,A^{p\Lambda^0}_{K^-pp} \{[\bar{u}_{\Lambda^0}(\vec{k}_{\Lambda^0},\sigma_{\Lambda^0})
  \gamma^5 u_p(\vec{k}_1,+\frac{1}{2})][\bar{u}_p(\vec{k}_p,\sigma_p) u_p(\vec{k}_2,-\frac{1}{2})] 
  \nonumber\\
  \hspace{-0.3in}&&
  - [\bar{u}_{\Lambda^0}(\vec{k}_{\Lambda^0},\sigma_{\Lambda^0})
  \gamma^5 u_p(\vec{k}_2,- \frac{1}{2})][\bar{u}_p(\vec{k}_p,\sigma_p) u_p(\vec{k}_1,
  + \frac{1}{2})]\}+ i \,B^{p\Lambda^0}_{K^-pp} \{[\bar{u}_{\Lambda^0}(\vec{k}_{\Lambda^0},\sigma_{\Lambda^0})
  u_p(\vec{k}_1,+ \frac{1}{2})]\nonumber\\
  \hspace{-0.3in}&&[\bar{u}_p(\vec{k}_p,\sigma_p)\gamma^5  u_p(\vec{k}_2,- \frac{1}{2})]
  -\,[\bar{u}_{\Lambda^0}(\vec{k}_{\Lambda^0},\sigma_{\Lambda^0})
  u_p(\vec{k}_2,-\frac{1}{2})][\bar{u}_p(\vec{k}_p,\sigma_p)\gamma^5 u_p(\vec{k}_1,+ \frac{1}{2})]\}.
\end{eqnarray} 
Taking into account that the $pp$ pair is in the ${^1}{\rm S}_0$
state, due to Fierz transformation the amplitude (\ref{label6.1}) can
be transcribed into the more compact form
\begin{eqnarray}\label{label6.2}
  \hspace{-0.3in} M(K^-pp \to p\Lambda^0) =
  \,i\,C^{p\Lambda^0}_{K^-pp}\,[\bar{u}_{\Lambda^0}(\vec{k}_{\Lambda^0},\sigma_{\Lambda^0})u^c_p(\vec{k}_p,\sigma_p)]
  \,[ \bar{u^c}_p(\vec{k}_2,-\frac{1}{2})\gamma^5 u_p(\vec{k}_1,+\frac{1}{2})],
\end{eqnarray} 
where $C^{p\Lambda^0}_{K^-pp} = (- A^{p\Lambda^0}_{K^-pp} -
B^{p\Lambda^0}_{K^-pp})/2$.  The effective coupling constants
$A^{p\Lambda^0}_{K^-pp}$ and $B^{p\Lambda^0}_{K^-pp}$ are calculated
with the chiral Lagrangians in Appendix B and weighted with the wave
function of the KNC ${^2_{\bar{K}}}{\rm H}$. They are equal to
\begin{figure} \centering
\psfrag{K-}{$K^-$} 
\psfrag{L0}{$\Lambda^0$} \psfrag{p0}{$\pi^0,\eta$}
\psfrag{p}{$p$}
 \psfrag{a}{$+$}
\psfrag{b}{$+ ~\ldots $}
\includegraphics[height= 0.10\textheight]{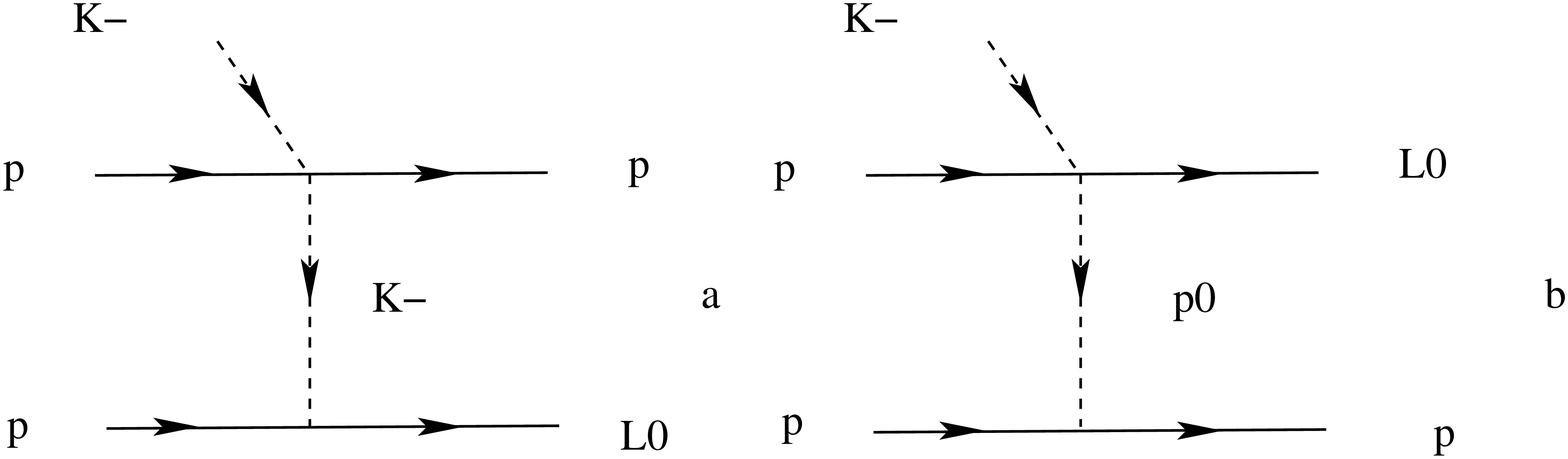}
\caption{The Feynman diagrams giving in the tree--approximation the
  main contribution to the amplitude of the reaction $K^- p p \to p
  \Lambda^0$ to leading order in the large $N_C$ and chiral
  expansion.}
\end{figure}
\begin{eqnarray}\label{label6.3}
  A^{p\Lambda^0}_{K^-pp} &=&  g_{\pi NN} \frac{3 - 2\alpha_D}{\sqrt{3}}
  \frac{m_K }{F^2_{\pi}}\,\frac{f_K(\omega)}{2 q_0k_0} = \,
  2.0\times 10^{-6}\,{\rm MeV}^{-3},\nonumber\\
  B^{p\Lambda^0}_{K^-pp} &=& g_{\pi NN} \frac{\sqrt{3}}{4}
  \frac{m_K}{F^2_{\pi}} \Big[\frac{f_{\pi}(\omega)}{2 q_0k_0}
  + (3 - 4\alpha_D) \frac{f_{\eta}(\omega)}{2 q_0k_0}\Big] = 2.0\times 10^{-6}\,{\rm MeV}^{-3},
\end{eqnarray} 
where $g_{\pi NN} = g_A m_N/F_{\pi} = 13.3$ is the $\pi NN$ coupling
constant \cite{PSI2,TE04}, the functions $f_K(\omega)$,
$f_{\pi}(\omega)$ and $f_{\eta}(\omega)$ are defined in Appendix C.
This gives $C^{p\Lambda^0}_{K^-pp} = -\,2.0\times 10^{-6}\,{\rm
  MeV}^{-3}$.
\begin{figure}
 \centering
\psfrag{K-}{$K^-$} 
\psfrag{L0}{$\Lambda^0$} \psfrag{p0}{$\pi^0,\eta$}
\psfrag{p}{$p$}
 \psfrag{a}{$+$}
\psfrag{Y}{$\Lambda^0,\Sigma^0$}
\psfrag{b}{$+ ~\ldots $}
\includegraphics[height= 0.12\textheight]{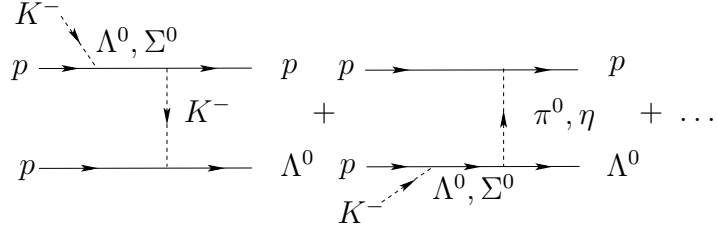}
\caption{The Feynman diagrams giving in the tree--approximation the
  main corrections to the effective coupling constant
  $C^{\,p\Lambda^0}_{K^-pp}$ of the transition $K^- p p \to p
  \Lambda^0$ to leading order in the large $N_C$ and chiral
  expansion.}
\end{figure}

The partial width of the decay ${^2_{\bar{K}}}{\rm H}\to p\Lambda^0$
(see ({\rm A}.2)) amounts to
\begin{eqnarray}\label{label6.4}
  \Gamma({^2_{\bar{K}}}{\rm H} \to p\Lambda^0) 
  =
  \frac{1}{\pi}
  \frac{|C^{p\Lambda^0}_{K^-pp}|^2\, k^3_0}{m_K(2 m_N + m_K +
    \epsilon_{{^2_{\bar{K}}}{\rm H}})}\Big|\int \frac{d^3q }{(2\pi)^3} \Phi_{K^-pp}(\vec{q}\,)\Big|^2
\Big|\int \frac{d^3Q}{(2\pi)^3}\Phi_{pp}(\vec{Q}\,)\Big|^2.
\end{eqnarray}
The integrals over the wave functions are calculated in
(\ref{label5.5}). This gives
\begin{eqnarray}\label{label6.5}
  \hspace{-0.3in} \Gamma({^2_{\bar{K}}}{\rm H} \to p\Lambda^0) 
  = \,|C^{p\Lambda^0}_{K^-pp}|^2\,\frac{m^{3/4}_N\mu^{9/4}
  }{2^{7/4}\pi^4 m_K} 
  \frac{
    k^3_0\,\omega^3}{2 m_N + m_K +
    \epsilon_{{^2_{\bar{K}}}{\rm H}}}= 34\,{\rm MeV}.
\end{eqnarray}
In the tree--approximation the correction $\delta
C^{\,p\Lambda^0}_{K^-pp}$to the effective coupling constant
$C^{\,p\Lambda^0}_{K^-pp}$ is given by the diagrams in Fig.\,5.  It is
of next--to--leading order in the large $N_C$ and chiral expansion.
For $N_C = 3$ it amounts to
\begin{eqnarray}\label{label6.6}
 \hspace{-0.3in} &&\delta C^{p\Lambda^0}_{K^-pp} = \frac{3 - 2\alpha_D}{4\sqrt{3}}
  \frac{ m^2_K}{m^2_N}\,\frac{g^3 _{\pi NN} }{2 m_N + m_K}
  \Big[\Big(\frac{3 - 2\alpha_D}{\sqrt{3}}\Big)^2 + (2\alpha_D - 1)^2\Big]
  \frac{f_K(\omega)}{4 q_0k_0}\nonumber\\
 \hspace{-0.3in} &&-\frac{\alpha_D}{\sqrt{3}}\,
\frac{3 - 2\alpha_D}{2\sqrt{3}}\frac{3 - 4\alpha_D}{\sqrt{3}}
  \frac{ m^2_K}{m^2_N}\,\frac{g^3 _{\pi NN} }{2 m_N + m_K} \frac{f_{\eta}(\omega)}{4 q_0k_0} 
  - \frac{\alpha_D}{\sqrt{3}}\,\frac{2\alpha_D - 1}{2}
  \frac{ m^2_K}{m^2_N}\,\frac{g^3 _{\pi NN} }{2 m_N + m_K} 
\frac{f_{\pi}(\omega)}{4 q_0k_0}=\nonumber\\
 \hspace{-0.3in} &&= 0.04\times 10^{-6}\,{\rm MeV}^{-3}.
\end{eqnarray} 
This makes up $2\,\%$ of the contribution given by the Feynman
diagrams in Fig.\,4.

The partial widths of the ${^2_{\bar{K}}}{\rm H} \to p \Sigma^0$ and
${^2_{\bar{K}}}{\rm H} \to n \Sigma^+$ decays can be calculated in the
analogous way.

\subsection{Total width of non\,--\,pionic decay channels}

The result of the calculation of the partial widths of the
${^2_{\bar{K}}}{\rm H} \to p \Sigma^0$ and ${^2_{\bar{K}}}{\rm H} \to
n \Sigma^+$ decays is
\begin{eqnarray}\label{label6.7}
  \Gamma({^2_{\bar{K}}}{\rm H} \to p\Sigma^0) = 2\,{\rm MeV}\quad,\quad
  \Gamma({^2_{\bar{K}}}{\rm H} \to n\Sigma^+) =  22\,{\rm MeV}.
\end{eqnarray}
The effective coupling constants of the transitions $K^-pp \to
p\Sigma^0$ and $K^-pp \to n\Sigma^+$  amount to
\begin{eqnarray}\label{label6.8}  
C^{p\Sigma^0}_{K^-pp} = -
      0.6\times 10^{-6}\,{\rm MeV}^{-3}\quad,\quad
C^{n\Sigma^+}_{K^-pp} = -  2.3\times 10^{-6}\,{\rm MeV}^{-3}.
\end{eqnarray}  
These coupling constants we use for the calculation of the
two\,--\,nucleon absorption contribution to the binding energy of the
KNC ${^2_{\bar{K}}}{\rm H}$.

The total width of the non\,--\,pionic decay channels
${^2_{\bar{K}}}{\rm H} \to NY$ with $NY = p\Lambda^0$, $p\Sigma^0$ and
$n\Sigma^+$, is equal to
\begin{eqnarray}\label{label6.9}
  && \Gamma^{(\rm non-pion.)}_{{^2_{\bar{K}}}{\rm H}} = 
  \sum_{NY}\Gamma({^2_{\bar{K}}}{\rm H} \to 
  NY) = 58\,{\rm MeV}.
\end{eqnarray}
This agrees well with the experimental data by the FINUDA
Collaboration $\Gamma^{(\exp)}_{{^2_{\bar{K}}}{\rm H}} =
(67^{+14}_{-11}(\rm stat)^{+2}_{-3}(syst))\,{\rm MeV}$ \cite{FINUDA}.

Using (\ref{label6.5}) and (\ref{label6.7}) we can calculate the ratio
\begin{eqnarray}\label{label6.10}
  R = \frac{\Gamma({^2_{\bar{K}}}{\rm H} \to 
    p\Lambda^0)}{\Gamma({^2_{\bar{K}}}{\rm H} \to 
    p\Sigma^0)} = \frac{I_{p\Lambda^0}}{I_{p\Sigma^0}} = 17,
\end{eqnarray}
where $I_{p\Lambda^0}$ and $I_{p\Sigma^0}$ are intensities of the
spectral lines in the invariant\,--\,mass spectra of the $p\Lambda^0$
and $p\Sigma^0$ pairs, respectively. Thus, this implies a negligible
small contribution of the ${^2_{\bar{K}}}{\rm H} \to p\Sigma^0 \to p
\Lambda^0 \gamma$ decay to the invariant--mass spectrum of the
$p\Lambda^0$ pair.

\section{Conclusion}
\setcounter{equation}{0}

We have proposed a phenomenological model for the KNC
${^2_{\bar{K}}}{\rm H}$, the binding energy and the width of which
agree well with the experimental data by the FINUDA Collaboration. The
binding energy and the width of the KNC ${^2_{\bar{K}}}{\rm H}$ have
been calculated in the form of momentum integrals of the real and
imaginary parts of the S\,--\,wave amplitude of elastic $K^-pp$
scattering weighted with the wave function of the KNC
${^2_{\bar{K}}}{\rm H}$ taking into account the main part of
correlations of the constituents. We have assumed that the KNC
${^2_{\bar{K}}}{\rm H}$ is the bound $K^-pp$ system, where the
$K^-$--meson is bound by the $pp$ pair in the ${^1}{\rm S}_0$ state.
For the analysis of the KNC ${^2_{\bar{K}}}{\rm H}$ we have used the
oscillator wave functions, describing the motion of the bound
$K^-$--meson relative to the $pp$ pair in the ${^1}{\rm S}_0$ state
and the relative motion of the protons in the $pp$ pair.  The
frequencies of oscillations are defined in terms of the frequency
$\omega = 64\,{\rm MeV}$, which we have fixed using the experimental
data on the mass and width of the strange baryon $\Lambda(1405)$.  We
have assumed that the $\Lambda(1405)$ is the KNC ${^1_{\bar{K}}}{\rm
  H}$.  Our assumption to treat the strange baryon $\Lambda(1405)$ as
a bound $K^-p$ isoscalar state with a width, caused by the $(K^-p)_{I
  = 0} \to \pi \Sigma$ decays, does not contradict the analysis of
this state in the coupled--channels techniques \cite{WW95,WW98}, where
the $\Lambda(1405)$ has been obtained as the $K^-p$ quasi--bound state
with isospin $I = 0$ and the resonance in the $\pi \Sigma$ channels.

The calculation of the real and imaginary parts of the S\,--\,wave
amplitude of elastic $K^-pp$ scattering we have carried out within the
chiral Lagrangian approach with the non--linear realization of chiral
$SU(3)\times SU(3)$ symmetry and derivative meson--baryon couplings,
describing strong low--energy interactions of ground baryon state 
octet and octet of pseudoscalar mesons. The amplitudes are calculated
in the tree--approximation and to leading order in the large $N_C$ and
chiral expansion, allowing to simplify the set of Feynman diagrams
contributing to the amplitudes of the reactions under consideration.

For the binding energy and the width of non\,--\,pionic decay channels
we have obtained $\epsilon_{{^2_{\bar{K}}}{\rm H}} = -\,118\,{\rm
  MeV}$ and $\Gamma^{(\rm non-pion.)}_{{^2_{\bar{K}}}{\rm H}} =
58\,{\rm MeV}$, respectively. They agree well with the experimental
data by the FINUDA Collaboration \cite{FINUDA}:
$\epsilon^{\exp}_{{^2_{\bar{K}}}{\rm H}} = (-\,115^{+6}_{-5})\,{\rm
  MeV}$ and $\Gamma^{(\exp)}_{{^2_{\bar{K}}}{\rm H}} =
(67^{+14}_{-11}(\rm stat)^{+2}_{-3}(syst))\,{\rm MeV}$. Since for the
binding energy $\epsilon_{{^2_{\bar{K}}}{\rm H}} = -\,118\,{\rm MeV}$
the pionic decay channels ${^2_{\bar{K}}}{\rm H} \to N\Sigma \pi$ are
suppressed kinematically and the decay channels ${^2_{\bar{K}}}{\rm H}
\to N\Lambda^0 \pi$ can be suppressed dynamically due to a possible
dominance of the intermediate $(K^-p)_{I = 0}$ state \cite{TDB1}, the
total width of non\,--\,pionic decay channels should be equal to the
total width of the KNC ${^2_{\bar{K}}}{\rm H}$ in the ground state
$\Gamma_{{^2_{\bar{K}}}{\rm H}} = 58\,{\rm MeV}$. For all that the
dominant decay channel is ${^2_{\bar{K}}}{\rm H} \to p\Lambda^0$,
whereas the decay channel ${^2_{\bar{K}}}{\rm H} \to p\Sigma^0$ is
strongly suppressed.

We would like to emphasise that the KNC ${^2_{\bar{K}}}{\rm H}$ is a
dense system \cite{TDB1}. The density of the KNC ${^2_{\bar{K}}}{\rm
  H}$ at the center of mass frame we define as
\begin{eqnarray}\label{label7.1}
  n_{{^2_{\bar{K}}}{\rm H}}(0) = 2|\Psi_{K^-pp}(0)|^2 = 
  \Big(\frac{2\mu\omega}{\pi}\Big)^{3/2} = 0.26\,{\rm fm}^{-3},
\end{eqnarray}
where $\Psi_{K^-pp}(0)$ is the wave function of the $K^-$--meson
oscillation relative to the $pp$ pair in the coordinate
representation. It is seen that the density of the KNC
${^2_{\bar{K}}}{\rm H}$ is by a factor of 2 larger than the normal
nuclear density $n_0 = 0.14\,{\rm fm}^{-3}$. However, in our approach
the influence of the dense nuclear matter \cite{PK04} on the dynamics
and the parameters of the constituents of the KNC ${^2_{\bar{K}}}{\rm
  H}$ is not taken into account. We are planning to carry out such an
analysis in our further investigations using, for example, the
technique developed in \cite{WW1}.

We have analysed the contribution of the Coulomb interaction to the
binding energy of the KNC ${^2_{\bar{K}}}{\rm H}$. We have obtained
$\delta \epsilon^{\rm Coul}_{{^2_{\bar{K}}}{\rm H}} = -\,1.35\,{\rm
  MeV}$.  This means that the Coulomb interaction does not destabilise
the $K^-pp$ system.

\section{Discussion: Akaishi--Yamazaki KNC $K^-pp$ in our approach}
\setcounter{equation}{0}

We have proposed an oscillator model of the KNC $K^-pp$, which
describes well the experimental data by the FINUDA Collaboration
\cite{FINUDA}.  Nevertheless, at first glimpse there is still a
problem concerning the origin of the discrepancy between the results
obtained within our description of the KNC $K^-pp$ and the potential
model approach for the KNC $K^-pp$, proposed by Akaishi and Yamazaki
\cite{TDB1}.  Indeed, the binding energies
$\epsilon_{{^2_{\bar{K}}}{\rm H} } = - 118\,{\rm MeV}$ and
$\epsilon_{{^2_{\bar{K}}}{\rm H} } = - 48\,{\rm MeV}$ of the KNC
$K^-pp$, obtained in our oscillator model and the potential model
approach by Akaishi and Yamazaki \cite{TDB1}, differ by a factor 2.5.

In this section we clarify this problem and show that a disagreement
with the results obtained by Akaishi and Yamazaki for the KNC $K^-pp$
is caused by different structures of the KNC $K^-pp$ in the
Akaishi--Yamazaki approach and in our model.  Indeed, in our model the
KNC $K^-pp$ is the bound state $K^-(pp)_{{^1}{\rm S}_0}$ of the
$K^-$--meson relative to the $pp$ pair in the ${^1}{\rm S}_0$ state,
whereas, according to Akaishi and Yamazaki \cite{TDB1,Yamazaki}, the
formation of the KNC $K^-pp$ goes through the resonance
$\Lambda(1405)$ (or $\Lambda^*$ \cite{TDB1,Yamazaki}), having the
structure $(K^-p)_{I = 0}$ \cite{TDB1}, which couples to the proton
producing the KNC $K^-pp$ with the structure $(K^-p)_{I = 0}\otimes
p$.

Below we show that the KNC $K^-pp$ with the structure $(K^-p)_{I =
  0}\otimes p$ can be fairly described in our oscillator model.

\subsection{Wave function of the KNC $(K^-p)_{I = 0}\otimes p$}

In our model the wave function of the KNC $(K^-p)_{I = 0}\otimes p$
should be equal to
\begin{eqnarray}\label{label8.1}
  \Phi_{(K^-p)_{I = 0}\otimes p}(\vec{q},\vec{Q}\,) &=&
  \Big(\frac{4\pi}{\mu_{Kp}\Omega_{Kp}}\Big)^{3/4}\,
\Big(\frac{4\pi}{\mu_{(Kp)p}\,\Omega_{(Kp)p}}\Big)^{3/4}\nonumber\\
  &\times&
  \exp\Big(-\,\frac{\vec{q}^{\;2}}{2\mu_{(Kp)p}\,
\Omega_{(Kp)p}}-\,\frac{\vec{Q}^{\;2}}{2\mu_{Kp}\Omega_{Kp}}\Big).
\end{eqnarray} 
The frequencies $\Omega_{Kp} = \Omega = 49.7\,{\rm MeV}$ (see
Eq.(\ref{label4.1})) and $\Omega_{(Kp)p} = \omega\sqrt{\mu/2
  \mu_{(Kp)p}} = 37.6\,{\rm MeV}$, where $\mu_{(Kp)p} = (m_K +
m_p)m_p/(m_K + 2 m_p) = 568\,{\rm MeV}$ is the reduced mass of the
$(Kp)p$ system, describe the motion of the $K^-$--meson relative to
the proton in the $(K^-p)_{I = 0}$ pair and the $(K^-p)_{I = 0}$ pair
relative to the proton, respectively.

Following our approach, the calculation of the binding energy and
partial widths of the KNC $(K^-p)_{I = 0} \otimes p$ we carry out
using the chiral Lagrangian with chiral $SU(3)\times SU(3)$ symmetry
and derivative meson--baryon couplings \cite{BWL68,JG83} (see Appendix
B), the large $N_C$ \cite{EW79}--\cite{AI1} and chiral expansion
\cite{JG83}.

\subsection{Binding energy of the KNC  $(K^-p)_{I = 0} \otimes p$}

The binding energy of the KNC $K^-pp = (K^-p)_{I = 0} \otimes p$ is
defined by the Weinberg--Tomozawa term of $(K^-p)_{I = 0}p$
scattering. This gives
\begin{eqnarray}\label{label8.2}
 -\, \epsilon_{(K^-p)_{I = 0}\otimes p} &=&\frac{3}{4}\,\frac{1}{F^2_{\pi}}\,
\Big(\frac{\mu_{Kp}\Omega_{Kp}}{\pi}\Big)^{3/2} 
 +\,\frac{1}{2}\,\frac{1}{F^2_{\pi}}\,\Big(\frac{\mu_{(Kp)p}\Omega_{(Kp)p}}{\pi}\Big)^{3/2} =\nonumber\\
  &=& (32 +\, 33)\,{\rm MeV} = 65\,{\rm MeV}.
\end{eqnarray} 
We would like to emphasize that Akaishi and Yamazaki normalised the
parameters of the KNC $K^-pp$ on the width of the $\Lambda*$ hyperon
equal to $\Gamma_{\Lambda^*} = 40\,{\rm MeV}$. In our model this gives
$\omega = 53\,{\rm MeV}$. In this case the binding energy of the KNC
$(K^-p)_{I = 0}\otimes p$ is $\epsilon_{(K^-p)_{I = 0}\otimes p} =
-\,49\,{\rm MeV}$ and agrees well with $\epsilon_{(K^-p)_{I =
    0}\otimes p} = -\,48\,{\rm MeV}$ by Akaishi and Yamazaki
\cite{TDB1,Yamazaki}.

For the binding energy $\epsilon_{(K^-p)_{I = 0}\otimes p} =
-\,65\,{\rm MeV}$ as well as $\epsilon_{(K^-p)_{I = 0}\otimes p} =
-\,49 \,{\rm MeV}$ kinematically allowed both non--pionic decay modes
$(K^-p)_{I = 0} \otimes p \to NY$, where $NY = p\Lambda^0, p\Sigma^0$
and $n\Sigma^+$, and pionic decay modes $(K^-p)_{I = 0} \otimes p \to
N\Sigma \pi$. The decay modes $(K^-p)_{I = 0} \otimes p \to N\Lambda^0
\pi$ are suppressed by the Akaishi--Yamazaki selection rules
\cite{TDB1,Yamazaki}.

\subsection{Non--pionic decay modes of the KNC $(K^-p)_{I = 0}\otimes p$}

The calculation of the partial widths of the non--pionic modes runs
parallel to that we have carried out in Section 6. This gives
$\Gamma^{p\Lambda^0}_{(K^-p)_{I = 0}\otimes p} = 19\,{\rm MeV}$ and
$\Gamma^{n\Sigma^+}_{(K^-p)_{I = 0}\otimes p} = 7\,{\rm MeV}$ for $\omega =
64\,{\rm MeV}$ and $\Gamma^{p\Lambda^0}_{(K^-p)_{I = 0}\otimes p} = 11\,{\rm MeV}$
and $\Gamma^{n\Sigma^+}_{(K^-p)_{I = 0}\otimes p}  = 4\,{\rm MeV}$ for $\omega
= 53\,{\rm MeV}$. The contribution of the $(K^-p)_{I = 0}p \to
p\Sigma^0$ decay mode is negligible small. Thus, the total partial
width of non--pionic decay modes is equal to
\begin{eqnarray}\label{label8.3}
 \Gamma^{\rm non-pionic}_{(K^-p)_{I = 0}\otimes p } = \left\{\begin{array}{r@{\quad,\quad}l}
26\,{\rm MeV}& \omega = 64\,{\rm MeV}\\
14\,{\rm MeV}& \omega = 53\,{\rm MeV}
\end{array}\right..
\end{eqnarray}
The width of non--pionic decays $\Gamma^{\rm non-pionic}_{(K^-p)_{I =
    0}\otimes p } = 14\,{\rm MeV}$ agrees well with the estimate
$\Gamma^{\rm non-pionic}_{(K^-p)_{I = 0}\otimes p } \approx 12\,{\rm
  MeV}$ obtained by Akaishi and Yamazaki \cite{TDB1}. 

\subsection{Pionic decay modes of the KNC $(K^-p)_{I = 0}\otimes p$}

For the binding energy $\epsilon_{K^-pp} = - 65\,{\rm MeV}$,
calculated for the width of the $\Lambda^*$ resonance
$\Gamma_{\Lambda^*} = 53\,{\rm MeV}$, as well as $\epsilon_{K^-pp} = -
49\,{\rm MeV}$, calculated for $\Gamma_{\Lambda^*} = 40\,{\rm MeV}$,
the pionic decay modes $(K^-p)_{I = 0}\otimes p \to NY\pi$ are opened
kinematically. Taking into account the Akaishi--Yamazaki selection
rule \cite{TDB1,Yamazaki}, the allowed decay modes are $(K^-p)_{I =
  0}\otimes p \to p\Sigma\pi $ and $(K^-p)_{I = 0}\otimes p \to
n\Sigma\pi$.

Since due to the $\Lambda^*$ doorway dominance \cite{Yamazaki}, the
main contribution comes from the decay modes $(K^-p)_{I = 0}\otimes p
\to p\Sigma\pi $, we calculate below only the partial widths of these
decay modes.  Feynman diagrams of the reactions $(K^-p)_{I = 0}\otimes
p \to p\Sigma\pi $ are depicted in Fig.\,6.
\begin{figure}
{\hspace{-1in}\centering 
\psfrag{K-}{$K^-$} 
\psfrag{L0}{$\Lambda^*$} 
\psfrag{S0}{$\Sigma$} 
\psfrag{p}{$p$}
\psfrag{pi}{$\pi$}
\psfrag{b}{$+ ~\ldots$}
\includegraphics[height= 0.15\textheight]{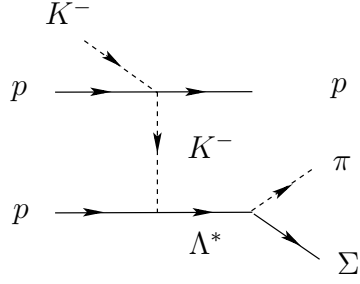}
\caption{The Feynman diagrams giving in the tree--approximation the
  main contribution to amplitude of the reactions $(K^-p)_{I =
    0}\otimes p \to p\Sigma \pi$ to leading order in the large $N_C$
  and chiral expansion.}}
\end{figure}

For the calculation of the amplitudes of the reactions $(K^-p)_{I =
  0}\otimes p \to p\Sigma \pi$ we use the chiral Lagrangian with
chiral $SU(3)\times SU(3)$ symmetry and derivative meson--baryon
couplings (see Appendix B). The coupling constants of the interaction
of the $\Lambda^*$ resonance with the ground baryon state octet and
octet of pseudoscalar mesons have been calculated in \cite{IV3}. The
partial width of the decay modes $(K^-p)_{I = 0}\otimes p \to p \Sigma
\pi$, defined by the diagrams in Fig.\,6, is equal to $\Gamma^{ p
  \Sigma \pi}_{(K^-p)_{I = 0}\otimes p} = 18\,{\rm MeV}$ and $\Gamma^{
  p \Sigma \pi}_{(K^-p)_{I = 0}\otimes p} = 27\,{\rm MeV}$ for the
frequencies $\omega = 64\,{\rm MeV}$ and $\omega = 53\,{\rm MeV}$,
respectively, where we have summed up over the decay modes $(K^-p)_{I
  = 0}\otimes p \to p \Sigma^+ \pi^-$, $(K^-p)_{I = 0}\otimes p \to p
\Sigma^- \pi^+$ and $(K^-p)_{I = 0}\otimes p \to p \Sigma^0 \pi^0$.

\subsection{Width of the KNC $(K^-p)_{I = 0}\otimes p$}

Summing up the contributions of the non--pionic and pionic decay modes
we get the value of the total width of the KNC with the structure
$(K^-p)_{I = 0}\otimes p$, proposed by Akaishi and Yamazaki
\cite{TDB1,Yamazaki}. The result is
\begin{eqnarray}\label{label8.4}
  \hspace{-0.3in} &&\Gamma_{(K^-p)_{I = 0}\otimes p} = 
\left\{\begin{array}{r@{\quad,\quad}l}
(44\pm 8)\,{\rm MeV}\,& \omega = 64\,{\rm MeV}\\
(41\pm 8)\,{\rm MeV}\,& \omega = 53\,{\rm MeV}
\end{array}\right..
\end{eqnarray}
The value of the total width $\Gamma_{(K^-p)_{I = 0}\otimes p} \approx
44\,{\rm MeV}$ is in qualitative agreement with that
$\Gamma_{(K^-p)_{I = 0}\otimes p} = 60\,{\rm MeV}$, predicted by
Akaishi and Yamazaki within the potential model approach. Such an
agreement can be made better taking into account the contributions of
the decay modes $(K^-p)_{I = 0}\otimes p \to n\Sigma\pi$.

\subsection{Summary}

The obtained results show that there is no discrepancy between our
approach and the potential model approach developed by Akaishi and
Yamazaki \cite{TDB1,Yamazaki}. Our model describes well both the
experimental data by the FINUDA Collaboration for the KNC $K^-pp$ with
the structure $K^-(pp)_{{^1}{\rm S}_0}$ and the Akaishi--Yamazaki
$K^-pp$ system with the structure $(K^-p)_{I = 0}\otimes p$ being the
consequence of the $\Lambda^*$ doorway dominance assumed by Yamazaki
and Akaishi \cite{Yamazaki} for the formation of the KNC $K^-pp$
\cite{TDB1}.

However, unlike the potential model approach by Akaishi and Yamazaki
\cite{TDB1}, which fails with the calculation of the partial widths of
non--pionic decay modes, our model, where the amplitudes of elastic
and inelastic of $K^-pp$ scattering are calculated using the chiral
Lagrangian with derivative meson--baryon couplings invariant under
non--linear realisation of chiral $SU(3)\times SU(3)$ symmetry, admits
the calculation of the partial widths of both non--pionic and pionic
decay modes.

\newpage

\section*{Appendix A: The binding energy and the width of the KNC
  ${^2_{\bar{K}}}{\rm H}$. The general formulas}
\renewcommand{\theequation}{A-\arabic{equation}}
\setcounter{equation}{0}

In our analysis of the KNC ${^2_{\bar{K}}}{\rm H}$ we propose to
describe the binding energy $\epsilon_{{^2_{\bar{K}}}{\rm H}}$ and the
width $\Gamma_{{^2_{\bar{K}}}{\rm H}}$ of the KNC ${^2_{\bar{K}}}{\rm
  H}$ by analogy to the energy level shift and width of the ground
state of kaonic deuterium \cite{IV3}. In this case
$\epsilon_{{^2_{\bar{K}}}{\rm H}}$ and $\Gamma_{{^2_{\bar{K}}}{\rm
    H}}$ can be defined in terms of the S\,--\,wave amplitude $M(K^-pp \to
K^-pp)$ of elastic $K^-pp$ scattering with the $pp$ pair in the
${^1}{\rm S}_0$ state weighted with the wave functions of the KNC
${^2_{\bar{K}}}{\rm H}$. We get
\begin{eqnarray}\label{labelA.1}
 &&  -\,\epsilon_{{^2_{\bar{K}}}{\rm H}} +\,i\,
  \frac{\Gamma_{{^2_{\bar{K}}}{\rm H}}}{2} =\nonumber\\
&& = \int \frac{d^3k }{(2\pi)^3}\,\frac{d^3K}{(2\pi)^3}
  \int \frac{d^3q }{(2\pi)^3}\,\frac{d^3Q}{(2\pi)^3}\,
  \frac{\Phi^*_{K^-pp}(\vec{k}\,)\Phi^*_{pp}(\vec{K}\,)\,
    \Phi_{K^-pp}(\vec{q}\,)\Phi_{pp}(\vec{Q}\,)}{\sqrt{2 E_K( \vec{k}\,) 2E_p(\vec{k}_1) 
      2E_p(\vec{k}_2) 2 E_K( \vec{q}\,) 2E_p(\vec{q}_1) 
      2E_p(\vec{q}_2)}}\nonumber\\
&&  \times\sum_{(\sigma_1,\sigma_2; {^1}{\rm S}_0)}M(K^-(\vec{q}\,) 
  p(\vec{q}_1,\sigma_1)p(\vec{q}_2,\sigma_2) \to K^-(\vec{k}\,) 
  p(\vec{k}_1,\sigma_1)p(\vec{k}_2,\sigma_2)),
\end{eqnarray}
where $\Phi_{K^-pp}(\vec{k}\,)$ is the wave function of a motion of
the $K^-$\,--\,meson relative to the $pp$ pair, and
$\Phi_{pp}(\vec{K}\,)$ is the wave function of the relative motion of
the protons in the $pp$ pair. Then, $\vec{q}$ and $\vec{k}$ are
relative momenta of the $K^-$\,--\,meson and the $pp$ pair in the
initial and the final states of the $K^-pp$ system, $\vec{Q}$ and
$\vec{K}$ are relative momenta of the protons in the $pp$ pair in the initial and the
final state of the $K^-pp$ system, $\vec{q}_1 = \vec{Q} - \vec{q}/2$
and $\vec{q}_2 = -\,\vec{Q} - \vec{q}/2$ are momenta of the protons in
the initial state, whereas $\vec{k}_1 = \vec{K} - \vec{k}/2$ and
$\vec{k}_2 = -\,\vec{K} - \vec{k}/2$ are momenta of the protons in the
final state.

The binding energy $\epsilon_{{^2_{\bar{K}}}{\rm H}}$ and the width
$\Gamma_{{^2_{\bar{K}}}{\rm H}}$ are defined by the real and imaginary
parts of the S\,--\,wave amplitude $M(K^-pp \to K^-pp)$, respectively.
The width $\Gamma_{{^2_{\bar{K}}}{\rm H}}$ is caused by the
two\,--\,body decays $K^-pp \to NY$, where $NY = p\Lambda^0,
p\Sigma^0$ and $n\Sigma^+$, and three\,--\,body decays $K^-pp \to
NY\pi$, where $NY\pi = N\Sigma\pi$ and $N\Lambda^0\pi$:
\begin{eqnarray}\label{labelA.2}
&&\Gamma_{{^2_{\bar{K}}}{\rm H}} = 
  \sum_{NY}\Gamma({^2_{\bar{K}}}{\rm H} \to NY) + \sum_{NY\pi}
  \Gamma({^2_{\bar{K}}}{\rm H} \to NY\pi ) = \nonumber\\
&&=  \frac{1}{16\pi^2}
  \sum_{NY}\int  \frac{d^3k_N}{ 
    E_N(\vec{k}_N)} \frac{d^3k_Y}{ 
    E_Y(\vec{k}_Y)}\,\delta^{(4)}(k_N + k_Y  
  - 2m_N - m_K - \epsilon_{{^2_{\bar{K}}}{\rm H}})\nonumber\\
&&\times\,\overline{\Big|\int \frac{d^3q }{(2\pi)^3}\,\frac{d^3Q}{(2\pi)^3}\,
    \frac{\Phi_{K^-pp}(\vec{q}\,)\Phi_{pp}(\vec{Q}\,)}{
      \sqrt{2 E_K( \vec{q}\,) 2E_p(\vec{q}_1) 
        2E_p( \vec{q}_2)}}\,M(K^-pp \to NY)\Big|^2}\nonumber\\
&&+ \frac{1}{256\pi^5}\sum_{NY\pi}
  \int \frac{d^3k_N}{ 
    E_N(\vec{k}_N)} \frac{d^3k_Y}{ 
    E_Y(\vec{k}_Y)} \frac{d^3k_{\pi}}{ 
    E_{\pi}(\vec{k}_{\pi})}\,\delta^{(4)}(k_N + k_Y + k_{\pi} 
  - 2m_N - m_K - \epsilon_{{^2_{\bar{K}}}{\rm H}})\nonumber\\
&&\times\,\overline{\Big|\int \frac{d^3q }{(2\pi)^3}\,\frac{d^3Q}{(2\pi)^3}\,
    \frac{\Phi_{K^-pp}(\vec{q}\,)\Phi_{pp}(\vec{Q}\,)}{
      \sqrt{2 E_K( \vec{q}\,) 2E_p( \vec{q}_1) 
        2E_p(\vec{q}_2)}}\,M(K^-pp \to NY\pi)\Big|^2}, 
\end{eqnarray}
where we have denoted
\begin{eqnarray}\label{labelA.3}
  &&\overline{\Big|\int \frac{d^3q }{(2\pi)^3}\,\frac{d^3Q}{(2\pi)^3}\,
    \frac{\Phi_{K^-pp}(\vec{q}\,)\Phi_{pp}(\vec{Q}\,)}{
      \sqrt{2 E_K( \vec{q}\,) 2E_p(\vec{q}_1) 
        2E_p( \vec{q}_2)}}\,M(K^-pp \to NY)\Big|^2} =\nonumber\\
&&= \frac{1}{4}\sum_{\sigma_1 = 
    \pm 1/2}\sum_{\sigma_2 = 
    \pm 1/2}\sum_{\alpha_1  = 
    \pm 1/2}\sum_{\alpha_2 = 
    \pm 1/2}\Big|\int \frac{d^3q }{(2\pi)^3}\,\frac{d^3Q}{(2\pi)^3}\,
  \frac{\Phi_{K^-pp}(\vec{q}\,)\Phi_{pp}(\vec{Q}\,)}{
    \sqrt{2 E_K( \vec{q}\,) 2E_p(\vec{q}_1) 
      2E_p( \vec{q}_2)}}\nonumber\\
&&\times
  M(K^-(\vec{q}\,)p(\vec{q}_1,\sigma_1)p(\vec{q}_2,\sigma_2) \to 
  N(\vec{k}_N,\alpha_1)Y(\vec{k}_Y,\alpha_2))\Big|^2,
\end{eqnarray}
and 
\begin{eqnarray}\label{labelA.4}
&&\overline{\Big|\int \frac{d^3q }{(2\pi)^3}\,\frac{d^3Q}{(2\pi)^3}\,
    \frac{\Phi_{K^-pp}(\vec{q}\,)\Phi_{pp}(\vec{Q}\,)}{
      \sqrt{2 E_K( \vec{q}\,) 2E_p( \vec{q}_1) 
        2E_p(\vec{q}_2)}}
    M(K^-pp \to 
    NY\pi)\Big|^2} =\nonumber\\
&&= \frac{1}{4}\sum_{\sigma_1 = 
    \pm 1/2}\sum_{\sigma_2 = 
    \pm 1/2}\sum_{\alpha_1  = 
    \pm 1/2}\sum_{\alpha_2 = 
    \pm 1/2}\Big|\int \frac{d^3q }{(2\pi)^3}\,\frac{d^3Q}{(2\pi)^3}\,
  \frac{\Phi_{K^-pp}(\vec{q}\,)\Phi_{pp}(\vec{Q}\,)}{
    \sqrt{2 E_K( \vec{q}\,) 2E_p( \vec{q}_1) 
      2E_p(\vec{q}_2)}}\nonumber\\
&&\times
  M(K^-(\vec{q}\,)p(\vec{q}_1,\sigma_1)p(\vec{q}_2,\sigma_2) \to 
  N(\vec{k}_N,\alpha_1)Y(\vec{k}_Y,\alpha_2)\pi(\vec{k}_{\pi}))\Big|^2.
\end{eqnarray}
We apply these formulas to the calculation of the binding energy and
the partial widths of the KNC ${^2_{\bar{K}}}{\rm H}$, caused by
non\,--\,pionic decay channels ${^2_{\bar{K}}}{\rm H} \to NY$.

\section*{Appendix B: Chiral Lagrangian for ground baryons coupled to
  pseudoscalar mesons within non--linear realization of chiral
  $SU(3)\times SU(3)$ symmetry}
\renewcommand{\theequation}{B-\arabic{equation}}
\setcounter{equation}{0}

For the calculation of amplitudes of elastic and inelastic $K^-pp$
scattering we use the chiral Lagrangian for the ground baryon state
octet, coupled to the pseudoscalar meson octet within the non--linear
realization of chiral $SU(3)\times SU(3)$ symmetry.  According to Lee
\cite{BWL68}, such a Lagrangian takes the form
\begin{eqnarray}\label{labelB.1}
&&{\cal L}_{\rm int}[B(x),P(x)] = {\rm
  tr}\{\bar{B}(x)i\gamma^{\mu}[s_{\mu}(x),B(x)]\}\nonumber\\
&&
 -\,g_A (1 - \alpha_D)\,{\rm tr}\{
\bar{B}(x)\gamma^{\mu}[p_{\mu}(x),B(x)]\} -\,g_A \alpha_D\,
{\rm tr}\{\bar{B}(x)\gamma^{\mu}\{p_{\mu}(x),B(x)\}\},
\end{eqnarray}
where the traces are calculated over $SU(3)$ indices, $g_A = 1.270$
and $\alpha_D = D/(D + F) = 0.635$ \cite{PDG04}. The Lagrangian (\ref{labelB.1}) 
is also used in ChPT for the description of strong low--energy
interactions of ground baryons with pseudoscalar mesons \cite{JG83}.

The fields of the octets of the ground baryons $\bar{B}(x)$ and $B(x)$
are defined by
\begin{eqnarray}\label{labelB.2}
\hspace{-0.3in}\bar{B}^b_a &=&\left(\begin{array}{llcl} {\displaystyle
\frac{\bar{\Sigma}^0}{\sqrt{2}} + \frac{\bar{\Lambda}^0}{\sqrt{6}}} &
\hspace{0.3in}\bar{\Sigma}^- & \bar{\Xi}^- \\
\hspace{0.3in}\bar{\Sigma}^+ &{\displaystyle -
\frac{\bar{\Sigma}^0}{\sqrt{2}} +
\frac{\bar{\Lambda}^0}{\sqrt{6}}} & \bar{\Xi}^0 \\
\hspace{0.3in}\bar{p}& \hspace{0.3in} \bar{n} & {\displaystyle
-\frac{2}{\sqrt{6}}\bar{\Lambda}^0} \\
\end{array}\right)\nonumber\\
B^b_a &=& \left(\begin{array}{llcl} {\displaystyle
\frac{\Sigma^0}{\sqrt{2}} + \frac{\Lambda^0}{\sqrt{6}}} &
\hspace{0.3in}\Sigma^+ & p \\ \hspace{0.3in}\Sigma^- &{\displaystyle -
\frac{\Sigma^0}{\sqrt{2}} + \frac{\Lambda^0}{\sqrt{6}}} & n \\
\hspace{0.3in}\Xi^- & \hspace{0.3in}\Xi^0 & {\displaystyle
-\frac{2}{\sqrt{6}}\Lambda^0}
\end{array}\right)\!,
\end{eqnarray}
where $\bar{B}^2_1 = \bar{\Sigma}^-$, $B^2_1 = \Sigma^+$ and so on. The
fields $s_{\mu}(x)$ and $p_{\mu}$ are defined by \cite{BWL68}
\begin{eqnarray}\label{labelB.3}
  s_{\mu}(x) &=& \frac{1}{2}\,(U^{\dagger}(x)\partial_{\mu}U(x) +
  U(x)\partial_{\mu}U^{\dagger}(x)),\nonumber\\
  p_{\mu}(x) &=&
  \frac{1}{2i}\,(U^{\dagger}(x)\partial_{\mu}U(x) -
  U(x)\partial_{\mu}U^{\dagger}(x)).
\end{eqnarray}
Here  $U(x)$ is a $3\times 3$ matrix given by
\begin{eqnarray}\label{labelB.4}
U^2(x) = e^{\textstyle\, \sqrt{2}\,i \gamma^5 P(x)/F_{\pi}}
\end{eqnarray}
where $F_{\pi} = 92.4\,{\rm MeV}$ is the PCAC constant of pseudoscalar
mesons and $P(x)$ is the octet of pseudoscalar mesons
\begin{eqnarray}\label{labelB.5}
P^a_b = \left(\begin{array}{llcl} {\displaystyle
\frac{\pi^0}{\sqrt{2}} + \frac{\eta}{\sqrt{6}}} &
\hspace{0.3in} \pi^+ & K^+\\ \hspace{0.3in} \pi^-
&{\displaystyle - \frac{\pi^0}{\sqrt{2}} +
\frac{\eta}{\sqrt{6}}} & K^0\\
\hspace{0.15in}- K^- & \hspace{0.3in} \bar{K}^0& {\displaystyle
-\frac{2}{\sqrt{6}}\,\eta} \\
\end{array}\right),
\end{eqnarray}
where $P^2_1 = \pi^+$, $P^3_1 = K^+$ and so on.  For simplicity we
identify the eighth component of the pseudoscalar octet $\eta(x)$ with
the observed pseudoscalar meson $\eta(550)$ \cite{JG83}.

For the practical application of the Lagrangian Eq.(\ref{labelB.1}) to the
calculation of the amplitudes of $K^-pp$ scattering there are useful
the following expansions
\begin{eqnarray}\label{labelB.6}
&&U^{\dagger}(x)\partial_{\mu}U(x) = +\,\gamma^5
\frac{ia}{F_{\pi}} P_{\mu}(x) + \frac{a^2}{2
  F^2_{\pi}} [P(x),P_{\mu}(x)] +
\gamma^5\frac{ia^3}{6F^3_{\pi}} [[P(x),P_{\mu}(x)],P(x)]\nonumber\\
&&
-\frac{a^4}{24 F^4_{\pi}}\,[[[P(x),P_{\mu}(x)],P(x)], P(x)] + \ldots,\nonumber\\
&&
U(x)\partial_{\mu}U^{\dagger}(x) = -\,\gamma^5
\frac{ia}{F_{\pi}} P_{\mu}(x) + \frac{a^2}{2
  F^2_{\pi}} [P(x),P_{\mu}(x)] -\,\gamma^5
\frac{ia^3}{6F^3_{\pi}} [[P(x),P_{\mu}(x)],P(x)]\nonumber\\
&&
-\frac{a^4}{24 F^4_{\pi}}\,[[[P(x),P_{\mu}(x)],P(x)], P(x)] +
\ldots,
\end{eqnarray}
where $a = 1/\sqrt{2}$ and $P_{\mu}(x) = \partial_{\mu}P(x)$. For
$s_{\mu}(x)$ and $p_{\mu}(x)$ we get the expansions
\begin{eqnarray}\label{labelB.7}
s_{\mu}(x) &=& \frac{a^2}{2 F^2_{\pi}}\,[P(x),P_{\mu}(x)] -
\,\frac{a^4}{24 F^4_{\pi}}\,[[[P(x),P_{\mu}(x)],P(x)], P(x)] +
\ldots,\nonumber\\
p_{\mu}(x) &=& \gamma^5 \frac{a}{F_{\pi}}\,P_{\mu}(x) + \gamma^5
\frac{a^3}{6F^3_{\pi}}\,[[P(x),P_{\mu}(x)],P(x)] + \ldots,
\end{eqnarray}
Since for $N_C \to \infty$ the PCAC constant $F_{\pi}$ behaves as
$F_{\pi} = O(\sqrt{N_C})$ \cite{EW79,HL00} (see also \cite{MR96}),
such an expansion corresponds to the large $N_C$ expansion in powers
of $1/N_C$ \cite{EW79,HL00,MR96}.

For the chiral Lagrangian of $\pi N$
interactions, invariant under chiral $SU(2)\times SU(2)$ symmetry, we
obtain \cite{SW67,SW68}
\begin{eqnarray}\label{labelB.8}
{\cal L}_{\pi N}(x) &=&-\,
\frac{g_A}{ F_{\pi}}\,\bar{N}(x)\gamma^{\mu}\gamma^5\,\frac{1}{2}\,\vec{\tau}\,N(x)\cdot
\partial_{\mu}\vec{\pi}(x) \nonumber\\
&&-\,\frac{1}{4 F^2_{\pi}}
\bar{N}(x)\gamma^{\mu}\vec{\tau}\,N(x)\cdot (\vec{\pi}(x) \times
\partial_{\mu}\vec{\pi}(x)) \ldots, 
\end{eqnarray}
where $N(x)$ is the nucleon field. It is the column matrix with elements
$(p(x),n(x))$, $\vec{\tau}$ are $2\times 2$ Pauli matrices and
$\vec{\pi}(x)$ is the $\pi$--meson field.

The chiral Lagrangian, applied to the calculation of the amplitudes of
$K^-pp$ scattering, is
\begin{eqnarray}\label{labelB.9}
&&{\cal L}_{\rm int}(x) =  {\cal L}_{PBB}(x) + {\cal L}_{PPBB}(x) = - g_A(1 -\alpha_D)\frac{a}{F_{\pi}}{\rm
  tr}\{\bar{B}(x)\gamma^{\mu}\gamma^5[P_{\mu}(x),B(x)]\} \nonumber\\ 
&&- g_A \alpha_D\frac{a}{F_{\pi}}\,{\rm
  tr}\{\bar{B}(x)\gamma^{\mu}\gamma^5\{P_{\mu}(x),B(x)\}\}+ \frac{a^2}{2 F^2_{\pi}}\,{\rm
  tr}\{\bar{B}(x)i\gamma^{\mu}[[P(x),P_{\mu}(x)], B(x)]\}\nonumber\\ 
&&+ \ldots
\end{eqnarray}
In the component form the interactions, defining the amplitudes of
$K^-pp$ scattering in the tree--approximation, read
\begin{eqnarray}\label{labelB.10}
  &&{\cal L}_{PBB} =
  -\,g_A\,\frac{a}{F_{\pi}}\,\{\bar{p}\gamma^{\mu}\gamma^5n\pi^+_{\mu} +
  \bar{n}\gamma^{\mu}\gamma^5p\pi^-_{\mu} +
  \frac{1}{\sqrt{2}}\,(\bar{p}\gamma^{\mu}\gamma^5p -
  \bar{n}\gamma^{\mu}\gamma^5n)\pi^0_{\mu}\nonumber\\
  &&
  +
  \frac{1}{\sqrt{6}}\,(3 - 4\alpha_D)\,(\bar{p}\gamma^{\mu}\gamma^5p +
  \bar{n}\gamma^{\mu}\gamma^5n)\eta_{\mu}
  +\frac{1}{\sqrt{6}}\,(3 - 2 \alpha_D)\,\bar{\Lambda}^0\gamma^{\mu}\gamma^5 p
  K^-_{\mu}\nonumber\\
  && - \frac{1}{\sqrt{2}}\,(2 \alpha_D - 1)\,\bar{\Sigma}^0\gamma^{\mu}\gamma^5 p
  K^-_{\mu} - (2 \alpha_D - 1)\,\bar{\Sigma}^-\gamma^{\mu}\gamma^5 n
  K^-_{\mu}
  - \frac{1}{\sqrt{6}}\,(3 - 2 \alpha_D)\,\bar{\Lambda}^0\gamma^{\mu}\gamma^5 n
  \bar{K}^0_{\mu}\nonumber\\
  && - \frac{1}{\sqrt{2}}\,(2 \alpha_D - 1)\,\bar{\Sigma}^0\gamma^{\mu}\gamma^5 n
  \bar{K}^0_{\mu} + (2 \alpha_D - 1)\,\bar{\Sigma}^+\gamma^{\mu}\gamma^5 p
  \bar{K}^0_{\mu}
  -\frac{1}{\sqrt{6}}\,(3 - 2 \alpha_D)\,\bar{p}\gamma^{\mu}\gamma^5 \Lambda^0
  K^+_{\mu} \nonumber\\
  &&
  + \frac{1}{\sqrt{2}}\,(2 \alpha_D - 1)\,\bar{p}\gamma^{\mu}\gamma^5 \Sigma^0
  K^+_{\mu} + (2 \alpha_D - 1)\,\bar{n}\gamma^{\mu}\gamma^5 \Sigma^-
  K^+_{\mu}
  -\frac{1}{\sqrt{6}}\,(3 - 2 \alpha_D)\,\bar{n}\gamma^{\mu}\gamma^5 \Lambda^0
  K^0_{\mu}\nonumber\\
  && - \frac{1}{\sqrt{2}}\,(2 \alpha_D - 1)\,\bar{n}\gamma^{\mu}\gamma^5 \Sigma^0
  K^0_{\mu} + (2 \alpha_D - 1)\,\bar{p}\gamma^{\mu}\gamma^5 \Sigma^+
  K^0_{\mu}
  +\sqrt{2}\,(1 - \alpha_D)\nonumber\\
  &&\times\,[\bar{\Sigma}^0\gamma^{\mu}\gamma^5(\Sigma^-
  \pi^+_{\mu} - \Sigma^+ \pi^-_{\mu}) +
  \bar{\Sigma}^-\gamma^{\mu}\gamma^5(\Sigma^0 \pi^-_{\mu} - \Sigma^-
  \pi^0_{\mu}) + \bar{\Sigma}^+\gamma^{\mu}\gamma^5(\Sigma^+ \pi^0_{\mu}
  - \Sigma^0 \pi^+_{\mu})]\nonumber\\
  &&
  + \alpha_D
  \sqrt{\frac{2}{3}}\,\bar{\Lambda}^0\gamma^{\mu}\gamma^5(\Sigma^0\pi^0_{\mu}
  + \Sigma^+ \pi^-_{\mu} + \Sigma^-\pi^+_{\mu}) + \alpha_D
  \sqrt{\frac{2}{3}}\,(\bar{\Sigma}^0\pi^0_{\mu} + \bar{\Sigma}^+
  \pi^+_{\mu} + \bar{\Sigma}^-\pi^-_{\mu})\gamma^{\mu}\gamma^5\Lambda^0\nonumber\\
  &&
  + \alpha_D \sqrt{\frac{2}{3}}\,(\bar{\Sigma}^+ \gamma^{\mu}\gamma^5
  \Sigma^+ + \bar{\Sigma}^- \gamma^{\mu}\gamma^5 \Sigma^- +
  \bar{\Sigma}^0 \gamma^{\mu}\gamma^5 \Sigma^0 - \bar{\Lambda}^0
  \gamma^{\mu}\gamma^5 \Lambda^0)\eta_{\mu} + \ldots
\end{eqnarray}
and 
\begin{eqnarray}\label{labelB.11}
  \hspace{-0.3in} &&{\cal L}_{PPBB} = i\,\frac{a^2}{2
    F^2_{\pi}}\,\Big\{\sqrt{2}\,\bar{p}\gamma^{\mu}n\,(\pi^0\pi^+_{\mu} -
  \pi^+ \pi^0_{\mu}) + \sqrt{2}\,\bar{n}\gamma^{\mu}p\,(\pi^- \pi^0_{\mu}
  - \pi^0 \pi^-_{\mu}) + (\bar{p}\gamma^{\mu}p -
  \bar{n}\gamma^{\mu}n) \nonumber\\
  \hspace{-0.3in} &&\times\,(\pi^+\pi^-_{\mu} - \pi^- \pi^+_{\mu}) 
  + \bar{p}\gamma^{\mu}p\,2\,(K^-K^+_{\mu} - K^+K^-_{\mu}) +
  \bar{n}\gamma^{\mu}n\,(K^-K^+_{\mu} - K^+K^-_{\mu}) \nonumber\\
  \hspace{-0.3in} &&
  + 
  \bar{n}\gamma^{\mu}p\,(K^- K^0_{\mu} - K^0 K^-_{\mu})
  - \bar{\Lambda}^0\gamma^{\mu}p\,\Big[\frac{\sqrt{3}}{2}\,(\pi^0
  K^-_{\mu} - K^- \pi^0_{\mu}) + \frac{3}{2}\,(\eta K^-_{\mu} - K^-
  \eta_{\mu})\Big]\nonumber\\
  \hspace{-0.3in} &&
  - \bar{\Sigma}^0\gamma^{\mu}p\,\Big[\frac{1}{2}\,(\pi^0 K^-_{\mu} -
  K^- \pi^0_{\mu}) + \,\frac{\sqrt{3}}{2}\,(\eta K^-_{\mu} - K^-
  \eta_{\mu})\Big] - \bar{\Sigma}^+\gamma^{\mu}p\,(\pi^+ K^-_{\mu} - K^-
  \pi^+_{\mu})\nonumber\\
  \hspace{-0.3in} && + \ldots\Big\}.
\end{eqnarray}
The first three terms in Eq.({\rm B}.11) reproduce the second term in
the Lagrangian ${\cal L}_{\pi N}$ Eq.({\rm B}.8).

In the non--linear realization of chiral symmetry with vector mesons
the low--energy interactions, defining the Weinberg--Tomozawa terms,
are obtained in the one--vector--meson exchange approximation
\cite{SW68}. The contribution of multiple--vector--meson exchanges are
of higher order in the large $N_C$ and chiral expansion.

\section*{Appendix C: Mesonic propagators weighted with the wave
  function of ${^2_{\bar{K}}}{\rm H}$}
\renewcommand{\theequation}{C-\arabic{equation}}
\setcounter{equation}{0}

For the calculation of the partial widths of the non--pionic
  decays ${^2_{\bar{K}}}{\rm H} \to NY$ of the KNC ${^2_{\bar{K}}}{\rm
    H}$, defined by the Feynman diagrams in Fig.\,4, we have to average
  also the mesonic propagators with the wave function of the KNC
  ${^2_{\bar{K}}}{\rm H}$:
\begin{eqnarray}\label{labelC.1}
\Big\langle \frac{1}{M^2 - Q^2_M}\Big\rangle = \int
\frac{d^3Q}{(2\pi)^3}\frac{d^3q}{(2\pi)^3}\,
\frac{\Phi_{pp}(\vec{Q}\,)\Phi_{K^-pp}(\vec{q}\,)}{\displaystyle M^2_{\rm eff}
  + (\vec{Q} + \frac{1}{2}\,\vec{q} -
  \,\vec{k}_0)^2},
\end{eqnarray}
where $M$ is the mass of the meson $K^-$, $\pi^-$ or $\eta$,
$\vec{k}_0$ is a relative momentum of the $NY$ pair and $M^2_{\rm eff}
= M^2 - (E_B - m_N)^2$, where $E_B$ is an energy of the daughter
baryon. This is a hyperon for the $K^-$--meson exchange and the
nucleon for the $\pi^0$ and $\eta$ meson exchanges.  The result of the
integration can be represented in the following form
\begin{eqnarray}\label{labelC.2}
\Big\langle \frac{1}{M^2 - Q^2_M}\Big\rangle = \frac{f_M(\omega)}{2 q_0k_0}\int
\frac{d^3Q}{(2\pi)^3}\,\Phi_{pp}(\vec{Q}\,) \int
\frac{d^3q}{(2\pi)^3}\,\Phi_{K^-pp}(\vec{q}\,).
\end{eqnarray}
The function $f_M(\omega)$ is defined by the integral
\begin{eqnarray}\label{labelC.3}
&&f_M(\omega) = 2q_0k_0 \int^{\infty}_0 \frac{dt}{\sqrt{1 +
    (m_N\Omega_{\parallel} + \mu \omega_{\parallel}/2)t}\,(1 +
  (m_N\Omega_{\parallel} + \mu \omega_{\perp}/2)t)}\nonumber\\
&&\times\,\exp\Big\{ - M^2_{\rm eff}t - \frac{k^2_0}{3}\,\Big(\frac{t}{1 + 
(m_N\Omega_{\parallel} +
    \mu \omega_{\parallel}/2)t} + \frac{2t}{1 + (m_N\Omega_{\parallel} +
    \mu \omega_{\perp}/2)t}\Big)\Big\},
\end{eqnarray}
where $q_0 = \sqrt{(1 + 2\sqrt{2}\,)\mu\omega/2} = 219\,{\rm MeV}$.

For the ${^2_{\bar{K}}}{\rm H} \to p \Lambda^0$ decay the functions
$f_M(\omega)$ are equal to: $f_K(\omega) = 0.50$, $f_{\pi}(\omega) =
1.00$ and $f_{\eta}(\omega) = 0.40$. The numerical values are obtained
for $\omega = 64\,{\rm MeV}$, $k_0 = 470\,{\rm MeV}$, $m_{\pi}
=135\,{\rm MeV}$ and $m_{\eta} = 548\,{\rm MeV}$. For the
${^2_{\bar{K}}}{\rm H} \to p \Sigma^0$ and ${^2_{\bar{K}}}{\rm H} \to
n \Sigma^+$ decays with the relative momentum $k_0 = 370\,{\rm MeV}$
of the $N \Sigma$ pair the functions $f_M(\omega)$ are equal to
$f_K(\omega) = 0.52$, $f_{\pi}(\omega) = 1.12$ and $f_{\eta}(\omega) =
0.35$.


\begin{thebibliography}{9}
\bibitem{IV3a} 
According to \cite{EW79}, the contribution of the
  exotic scalar mesons $a_0(980)$ and $f_0(980)$ with a four\,--\,quark
  structure $qq\bar{q}\bar{q}$ is of order $O(1/N^{3/2}_C)$ (see
  \cite{IV3}).
\end{thebibliography}
\end{document}